\documentclass[10 pt,final,journal,letterpaper,oneside,twocolumn]{IEEEtran}
\usepackage{multicol}   
\usepackage{multirow}
\usepackage[pdftex]{graphicx}
\usepackage[algo2e,ruled, vlined, linesnumbered]{algorithm2e}
\usepackage{algorithmic}
\usepackage[small]{caption}
\usepackage{float}
\usepackage{amssymb,amsmath}
\usepackage{graphicx}
\usepackage{subfigure}

\usepackage{xspace}
\usepackage{fancyhdr}
\usepackage{amssymb,amsmath}
\usepackage{textcomp} 
\usepackage{epstopdf}
\usepackage{bbding}
\usepackage{cite} 
\usepackage{color}
\usepackage{multicol}
\usepackage{multirow}
\usepackage{array}
\usepackage{pifont}
\usepackage{hyperref}
\begin{document}
\title{{Efficient Resource Allocation and User Association in NOMA-Enabled Vehicular-Aided HetNets with High Altitude Platforms}}
\author{Ali Nauman, Mashael Maashi, Hend K. Alkahtani, Fahd N. Al-Wesabi, Nojood O Aljehane,\\ Mohammed Assiri , Sara Saadeldeen Ibrahim, Wali Ullah Khan \thanks{ Ali Nauman is with the Department of Information and Communication Engineering, Yeungnam University, Republic of Korea.

Mashael Maashi is with the Department of Software Engineering, College of Computer and Information Sciences, King Saud University, Po box 103786, Riyadh 11543, Saudi Arabia.

Hend K. Alkahtani is with the Department of Information Systems, College of Computer and Information Sciences, Princess Nourah Bint Abdulrahman University, P.O. Box 84428, Riyadh 11671, Saudi Arabia.

Fahd N. Al-Wesabi is with the Department of Computer Science, College of Science \& Art at Mahayil, King Khalid University, Saudi Arabia.

Nojood O Aljehane is with the Department of Computer Science, Faculty of Computers and Information Technology, University of Tabuk, Tabuk, Saudi Arabia.

Mohammed Assiri is with the Department of Computer Science, College of Sciences and Humanities- Aflaj, Prince Sattam bin Abdulaziz University, Aflaj 16273, Saudi Arabia.

Sara Saadeldeen Ibrahim is with the Department of Computer and Self Development, Preparatory Year Deanship, Prince Sattam bin Abdulaziz University, AlKharj, Saudi Arabia.

Wali Ullah Kahn is with the Interdisciplinary Centre for Security, Reliability and Trust (SnT), University of Luxembourg, Luxembourg.

Corresponding Author:  Fahd N. Al-Wesabi, email: falwesabi@kku.edu.sa
}}

\markboth{Computer Communiaction}%
{Shell \MakeLowercase{\textit{et al.}}: Bare Demo of IEEEtran.cls for IEEE Journals} 


\maketitle
\begin{abstract}
The increasing demand for massive connectivity and high data rates has made the efficient use of existing spectrum resources an increasingly challenging problem. Non-orthogonal multiple access (NOMA) is a potential solution for future heterogeneous networks (HetNets) due to its high capacity and spectrum efficiency. In this study, we analyze an uplink NOMA-enabled vehicular-aided HetNet, where multiple vehicular user equipment (VUEs) share the access link spectrum, and a high-altitude platform (HAP) communicates with roadside units (RSUs) through a backhaul communication link. We propose an improved algorithm for user association that selects VUEs for HAPs based on channel coefficient ratios and terrestrial VUEs based on a caching-state backhaul communication link. The joint optimization problems aim to maximize a utility function that considers VUE transmission rates and cross-tier interference while meeting the constraints of backhaul transmission rates and QoS requirements of each VUE. The joint resource allocation optimization problem consists of three sub-problems: bandwidth allocation, user association, and transmission power allocation. We derive a closed-form solution for bandwidth allocation and solve the transmission power allocation sub-problem iteratively using Taylor expansion to transform a non-convex term into a convex one. Our proposed three-stage iterative algorithm for resource allocation integrates all three sub-problems and is shown to be effective through simulation results. Specifically, the results demonstrate that our solution achieves performance improvements over existing approaches.
\end{abstract}

\section{Introduction}
Integrated terrestrial non-terrestrial networks refer to telecommunications systems that effectively merge terrestrial and non-terrestrial communication infrastructure, hence offering complete and robust connectivity \cite{khan2023rate}. Integrating terrestrial networks, which possess advantageous characteristics such as high capacity and low latency, with non-terrestrial networks like satellite constellations and high-altitude platforms (HAPS) presents a versatile solution for addressing global communication requirements \cite{ren2023handoff}. This solution proves particularly beneficial in areas with limited terrestrial coverage, as well as in scenarios involving disaster recovery, Internet of Things (IoT) applications, aerospace, aviation, military operations, and emergency response \cite{ahmed2023survey}. By combining the strengths of both types of networks, this integrated approach ensures redundancy, high-speed data transmission, and dependable connectivity across diverse environments. With the increasing demand for connected devices and autonomous systems, future communication systems will rely heavily on integrating terrestrial and non-terrestrial networks to facilitate the delivery of cutting-edge services \cite{al2022next}. Together, they can provide more comprehensive and reliable communications, offering improved coverage, high spectrum efficiency, cost savings, and better use of resources \cite{dong2022intelligent}. Therefore, the convergence of terrestrial networks and satellites has garnered significant attention among researchers and professionals worldwide \cite{khan2022opportunities}.
\par
HAPs have received significant attention due to their ability to stay aloft at high altitudes for extended periods and provide various services such as communication, surveillance, and remote sensing \cite{kurt2021vision}. HAPs are unmanned aerial vehicles (UAVs) operating at high altitudes ranging from 12 to 22 kilometers (7.5 to 14 miles), above most weather patterns and commercial air traffic \cite{renga2022can}. HAPs have developed for several decades, but recent advances in materials, propulsion, and communication technologies have renewed interest in their potential applications. Additionally, HAPs can be deployed quickly and at a lower cost than satellites, making them an attractive alternative for certain applications \cite{liu2022interference}.
\par
Heterogeneous networks (also called HetNets), which combine terrestrial base stations (BSs) and non-terrestrial, i.e., UAVs, HAPs, and satellites networks, offer a range of benefits, including high-speed links for UEs provided by terrestrial BSs, global coverage and ample backhaul capacity by the non-terrestrial network \cite{feng2019enabling}. In terrestrial non-terrestrial HetNets, the challenges posed by the limited spectrum resource and the growing number of users who generate more interference must be addressed to ensure efficient transmission \cite{ji2021flexible}. Despite its ability to mitigate interference, orthogonal multiple access technology is subject to certain limitations, including restricted improvements in the utilization of spectrum and enhanced capacity because only one user is capable of connecting over a resource block (orthogonal) in a unit of time \cite{wang2020resource}.
\par
Moreover, integrating NOMA technologies \cite{khan2022energy} can significantly enhance the effective utilization of communication network spectrum resources \cite{khan2021joint}. Researchers widely recognize NOMA as a promising approach to allocate spectrum resources effectively and utilize them rationally in next-generation multiple access technologies \cite{ihsan2023energy}. NOMA facilitates the sharing of the same spectrum of resources among multiple users, leading to improved performance, particularly in non-terrestrial communication with extensive coverage \cite{asif2022energy}. Similarly, by incorporating NOMA into terrestrial satellite HetNets is gaining prominence as a promising approach to boost communication performance and optimize spectrum resource utilization \cite{zhao2022integrated}.
\subsection{Resent Academic Advances}
Numerous research efforts have aimed to enhance NOMA system performance, covering areas such as interference management, sum rate optimization, energy efficiency maximization, power and bandwidth allocation, and distributed cluster allocation. For example, in \cite{randriaerence} and \cite{budhi020cross}, authors proposed optimization algorithms to tackle the challenge of interference management in NOMA-based communication systems. In \cite{celikstributed}, researchers explored a scheme for optimizing power and bandwidth distribution in NOMA-based downlink heterogeneous networks, where each base station organizes into clusters and independently manages power and bandwidth allocation.
Additionally, in reference \cite{muhammed2020energy}, the authors delved into maximizing energy efficiency in NOMA-assisted networks through bandwidth and power allocation. Furthermore, they derived a mathematical closed-form equation and an optimization algorithm for power and bandwidth using the generalized Dinkelbach method. Several other research works have also investigated optimization problems in NOMA-assisted wireless networks \cite{ahmed2022cooperative, khan2022ambient, ihsan2022energy}.
In recent years, integrating NOMA with satellite communication systems has attracted considerable attention from researchers aiming to leverage the unique advantages of these technologies for optimal resource utilization. This trend has led to many studies, including those presented in \cite{wang2020resource, jiao2020network}, exploring various optimization techniques for NOMA-assisted satellite networks. For example, in \cite{wang2020resource}, a novel iterative method for User equipment association, subchannel allocation, and power allocation was proposed, resulting in improved system throughput compared to existing methods. In \cite{chu2020robust}, researchers focused on beamforming optimization in NOMA non-terrestrial IoT networks with a multi-beam architecture. Recent literature highlights the growing interest of researchers in optimizing resource utilization by integrating HAP communication systems with NOMA.
\par
Similarly, the authors of the referenced works \cite{zhu2017non, wang2019user} concentrate on designing beamforming vectors and efficient resource allocation in NOMA terrestrial and non-terrestrial systems, respectively. Additionally, in \cite{wang2019user}, authors addressed user association and power optimization by presenting a mathematical closed-form expression to find the power solution, which was then incorporated into user association schemes to achieve globally optimal results. Furthermore, reference \cite{jiao2020network} tackled the optimal allocation of resources in IoT-based NOMA-enabled terrestrial networks integrated with satellite communications. The authors proposed a heuristic algorithm, such as particle swarm optimization, for both power and bandwidth optimization, utilizing a Lyapunov framework to find the optimal solutions. This work underscores the potential of integrating NOMA with non-terrestrial networks to enhance communication performance in IoT terrestrial networks.
\par
The integration of wireless caching and NOMA has garnered substantial attention in recent years as researchers aim to enhance the performance and efficiency of wireless communication systems \cite{randriaerence}, \cite{zhang2020energy}-\cite{qi2019advanced}. Wireless caching, storing frequently accessed content at edge nodes in the network, can alleviate network congestion and enhance low-latency communication performance \cite{wang2014cache}, \cite{li2018survey}. This technology can also enhance the performance of terrestrial non-terrestrial HetNets by reducing the demand on the backhaul link \cite{song2021energy, li2020energy}.
Numerous studies have explored the integration of wireless caching and NOMA, examining various aspects of the technology and its potential advantages. In \cite{zhang2020energy}, researchers delved into optimizing energy-efficient resource allocation in NOMA networks incorporating terahertz communication and caching, intending to improve energy efficiency by judiciously utilizing resources in NOMA networks.
Additionally, in \cite{fu2019dynamic} and \cite{yang2020cache}, researchers investigated the application of deep learning algorithms for optimizing resource allocation in NOMA networks incorporating caching. In \cite{ding2018noma}, two NOMA-based caching strategies were presented to mitigate latency in content delivery. The challenge of user association and power allocation in NOMA networks incorporating caching was addressed in \cite{randriaerence} and \cite{qi2019advanced}, with the authors proposing joint algorithms to optimize these aspects. In conclusion, the integration of wireless caching and NOMA continues to be a subject of ongoing research and investigation.
\par
{\color{black}Recently, some researchers have also considered terrestrial-HAPS networks and evaluate various aspects of system performance. Alidadi {\em et al.} \cite{shamsabadi2023joint} have investigated a fairness problem in terrestrial-HAPS network to maximize the minimum spectral efficiency by optimizing the system resources. Ren {\em et al.} \cite{ren2023handoff} have provided an adaptive delay minimization problem and optimized task splitting, power control, spectrum assignment, and computational resource allocation. Moreover, the authors of \cite{kocc2023haps} have proposed a cell-switching
methods in terrestrial-HAPS integrated networks to improve the energy consumption of the system. Alfattani {\em et al.} \cite{alfattani2023resource} have proposed a new optimization framework to enhance the users connectivity and minimize the energy consumption in terrestrial-HAPS networks. Further, Zheng {\em et al.} \cite{zheng2023analysis} have studied the positioning performance of terrestrial-HAPS integrated networks to improve 3D positioning accuracy, the horizontal dilution of precision, and the vertical dilution of precision. Besides the above studies, Erdogan {\em et al.} \cite{erdogan2023optical} have proposed different use-cases of terrestrial-HAPS network and investigated the physical layer security of the system. In addition, Shafie {\em et al.} \cite{shafie2022power} have optimized the power allocation in terrestrial-HAPS network to maximize the spectral efficiency under special correlation of channel gain and imperfect NOMA signal decoding. Then, the work in \cite{cumali2022user} has provided user association and codebook design scheme for the spectral efficiency of terrestrial-HAPS integrated networks. Of late, authors have also studied interference issues \cite{shamsabadi2022handling}, mobile edge computing based task offloading \cite{ovatman2022accurate}, and terahertz communication in terrestrial-HAPS networks \cite{abbasi2022cell}.
\begin{table*}[t]
\label{T1}
	\centering
{\color{black}\caption{3GPP standardization works on integrated terrestrial non-terrestrial networks. \cite{lin20215g,lin2022overview}}
	\label{tab:simulation_parameters}
	\begin{tabular}{| m{1cm} | m{15cm}|}
\hline 
\hline
Release &      Advance in integrated  terrestrial non-terrestrial networks   \\
\hline
Rel-15 & The focus of Rel-15 is on New Radio (NR), a technology proposed for the purpose of supporting terrestrial non-terrestrial networks as outlined in the technical report [TR 38.811]. Additionally, this study provides pertinent use case possibilities for integrating terrestrial non-terrestrial networks and spectrum, explicitly focusing on the S-band and Ka-band frequencies. In addition, it delineates the dimensions of the footprint, the angle of assessment, the configuration of the beam, and the design of the antenna. Additionally, this release provides specific details regarding the channel propagation model as outlined in the technical report [TR 38.901].\\
\hline
Rel-16 & The present release presents potential resolutions for the integration of new technologies inside terrestrial non-terrestrial integrated networks, as outlined in the technical report [TR 38.821]. The primary emphasis is placed on utilizing FR1 bands inside terrestrial non-terrestrial networks to facilitate the seamless integration and functioning of the Internet of Things. Additionally, this approach facilitates the identification of necessary modifications in the physical layer and other layers, considering the assumptions made during system-level simulations. In addition to this, the present study also examines the influence of resource optimization on the performance of terrestrial non-terrestrial networks. Moreover, it integrates the utilization of terrestrial non-terrestrial networks in the context of next-generation communications, as indicated in the technical report [TR 22.822], to facilitate the provision of diverse services.
\\
\hline
Rel-17 & The topic of Rel-17 pertains to the inclusion of narrowband IoT and machine-type communication in integrated terrestrial non-terrestrial scenarios, as referenced in the technical report [TR 36.763]. The technology is primarily designed to meet the unique requirements of IoT applications. Considerable focus has been devoted to the architectural concerns for satellite access within the framework of 6G, as outlined in the technical report [TR 23.737]. This endeavor involves improvements in various aspects, such as advances in radio frequency and physical layer characteristics, optimizations in protocols, and more efficient management of radio resources. In addition, this process entails the selection of a suitable architectural framework, addressing challenges related to integrated terrestrial non-terrestrial roaming, and enhancing conditional handover procedures.
\\
\hline
Rel-18 & The advancements pertaining to terrestrial non-terrestrial communication will investigate the extent of system coverage for handheld devices in practical scenarios, as well as explore access capabilities beyond the 10 GHz frequency range for both stationary and mobile platforms. The study aims to investigate the necessary conditions for the network-validated user location and address challenges pertaining to user mobility and the uninterrupted provision of services during transitions between terrestrial and satellite networks, as well as various non-terrestrial networks.
\\
\hline
\end{tabular}   }
\end{table*}
\subsection{Recent Industry Advances}
The process of standardization for terrestrial non-terrestrial networks within the 3rd Generation Partnership Project (3GPP) commenced in the year 2017 \cite{lin20215g}. The standardization endeavor can be classified into two main domains: improvements for networks operating in non-terrestrial environments and enhancements for networks operating in terrestrial environments. The primary objective of this initiative is to develop a universally accepted benchmark for non-terrestrial communications, hence fostering substantial expansion within the satellite, HAPS, and UAV industry. The activities in the aforementioned domain have a twofold objective, which is to guarantee that mobile standards are in line with the connection needs for secure functioning on platforms that are not on Earth. Table II provides a comprehensive summary of the objectives and results achieved by 3GPP in its endeavors encompassing Rel-15 to Rel-17, together with the ongoing investigations for Rel-18.

In the context of 3GPP, terrestrial non-terrestrial networks pertain to the application of satellites or HAPS to provide connectivity services, specifically in geographically isolated regions where conventional cellular coverage is insufficient. The core set of features added by 3GPP in Rel-17 aims to enable next-generation spectrum operation over terrestrial-satellite networks within the frequency range of FR1, encompassing frequencies up to 7.125 GHz. The forthcoming Rel-18 of the 3GPP is focused on advancing the capabilities of next-generation operations inside terrestrial-satellite environments. The proposed enhancement aims to enhance the coverage of handheld devices, investigate the feasibility of deploying networks in frequency bands above 10 GHz, tackle challenges related to mobility, ensure uninterrupted service transition between terrestrial and non-terrestrial networks, and evaluate the regulatory obligations associated with verifying user locations within the network \cite{lin2022overview}.  

The inclusion of non-terrestrial platforms in the preceding network generation was initially included in 3GPP's Rel-15. This involved the incorporation of signaling protocols to identify extraterrestrial users using subscription-based techniques. Furthermore, protocols were implemented to facilitate the reporting of essential characteristics pertaining to non-terrestrial platforms, including height, position, speed, and flight trajectory. In order to efficiently address non-terrestrial interference, especially in situations where there is a specific concentration of low-altitude non-terrestrial platforms, novel measurement reports have been implemented for effective management.

In following iterations, the 3GPP expanded its scope to cater to the requirements of linked non-terrestrial systems at the application layer, with a significant emphasis on security considerations. These releases have also established the groundwork for establishing the protocols by which non-terrestrial platforms engage with the Traffic Management system, facilitating synchronized and secure operations of non-terrestrial platforms inside the network. The upcoming release of 3GPP's Rel-18 aims to provide specialized next-generation spectrum support that is specifically designed for devices running on aerial vehicles. This development is in response to the evolving use cases of next-generation technologies. According to \cite{lin2022overview}, the forthcoming progress will encompass the investigation of supplementary factors that can initiate conditional handover. Additionally, it will incorporate the utilization of base station uptilting approaches to increase communication, as well as the integration of signaling mechanisms to indicate the beamforming capabilities of non-terrestrial platforms, alongside various other improvements.
}
\subsection{Motivation and Contributions}
{\color{black}Resource allocation in wireless networks has been a topic of significant research interest. The focus has been improving the performance and efficiency of wireless caching networks, NOMA networks, and terrestrial-HAPs networks. Despite these efforts, most studies have only considered one network type in isolation rather than an integrated approach. Currently, there is a gap in the literature that addresses the comprehensive problem of efficient allocation of resources in caching-based NOMA for HAPs-terrestrial communication networks, particularly concerning bandwidth, power, and user association. The complexities of resource allocation in these networks arise from the requirement for reliable wireless backhaul access networks and the added challenge of interference in multi-cell NOMA terrestrial-satellite heterogeneous networks.

The objective of this study is to address the challenges presented by NOMA-enabled vehicular-aided HetNet, which involves RSU and HAPs VUEs sharing the same radio resource. The HAPs are responsible for back-haul communication links to the terrestrial RSU. To achieve this, the study begins by formulating a joint optimization problem considering the achievable rate for cellular VUEs and the interference from RSU-HAPs. An optimization problem is developed to consider factors such as user association, back-haul link limitations, and QoS requirements for each VUE. To further enhance the system performance, the joint optimization formulation is broken down into sub-problems, including user association, efficient allocation of bandwidth, and transmission power. Subsequently, algorithms are proposed to optimize the system performance. This paper provides a comprehensive analysis of the resource allocation process, including a time complexity analysis of the proposed optimization algorithms. The effectiveness of the proposed scheme is demonstrated through simulation results. The main contributions of this study include:}
\begin{enumerate}
    \item In this study, we present an up-link communication scenario in the HetNet model incorporating caching strategies for spectrum sharing between RSU and HAPs VUEs is presented. To ensure optimal system performance, the interplay of inter-cell, cross-tier, as well as intra-cell interference is analyzed. The optimal decoding order for successive interference cancellation (SIC) is then calculated. To assess system performance, a utility function is created that considers the achievable rate of cellular UEs and the cross-tier interference generated by RSUs towards the satellite. The resource allocation problem is then posed as a system utility maximization problem, with joint user associations, back-haul constraints, and VUE QoS requirements all considered.
    \item The resource allocation optimization problem in this system model is highly non-linear and computationally complex, with Non-deterministic Polynomial (NP) complexity. To tackle this challenge, the problem is decomposed into three sub-problems, which are solved independently. The first sub-problem focuses on the VUE-AP association and proposes an advanced preference relation, caching, and swapping-based algorithm. The algorithm's implementation involves prioritizing satellite VUEs based on the channel coefficient ratios and considering the caching state and availability of the back-haul communication link in the association of terrestrial VUEs. The second sub-problem yields a closed-form expression for bandwidth allocation. In contrast, the third sub-problem involves transforming the non-convex objective function optimization into a convex form through successive convex approximation, which is then solved iteratively through a power allocation algorithm.
    \item To optimize the resource allocation in the system, a three-stage iterative algorithm is proposed. The proposed approach for resolving the optimal allocation of resources in the up-link up-link communication scenario in HetNet communication involves the iterative solution of three sub-problems, specifically user association, bandwidth allocation, and power allocation. A detailed analysis of the time complexity of the proposed algorithms is carried out to evaluate their computational efficiency. The validity of the algorithms' efficacy in optimizing resource allocation is established through simulation results, which also demonstrate the convergence of the proposed algorithms.
\end{enumerate}
\section{Network Model \& Mathematical Formulation}
{\color{black} This section presents the caching-based NOMA-enabled vehicular-aided HetNet, which aims to enhance communication network performance by efficiently using network resources. To achieve this goal, a utility function is constructed, and an optimization problem is formulated, considering three essential elements of the network: bandwidth assignment for front-haul and back-haul communication, user association, and power allocation. This approach leverages the advantages of NOMA as compared to OMA, such as improved spectrum efficiency and flexible interference management, to optimize resource allocation in the context of vehicular-aided HetNets\cite{liu2022developing,liu2022evolution}.}

\begin{figure*}[!t]
	\centering
	\includegraphics[width=0.5\linewidth]{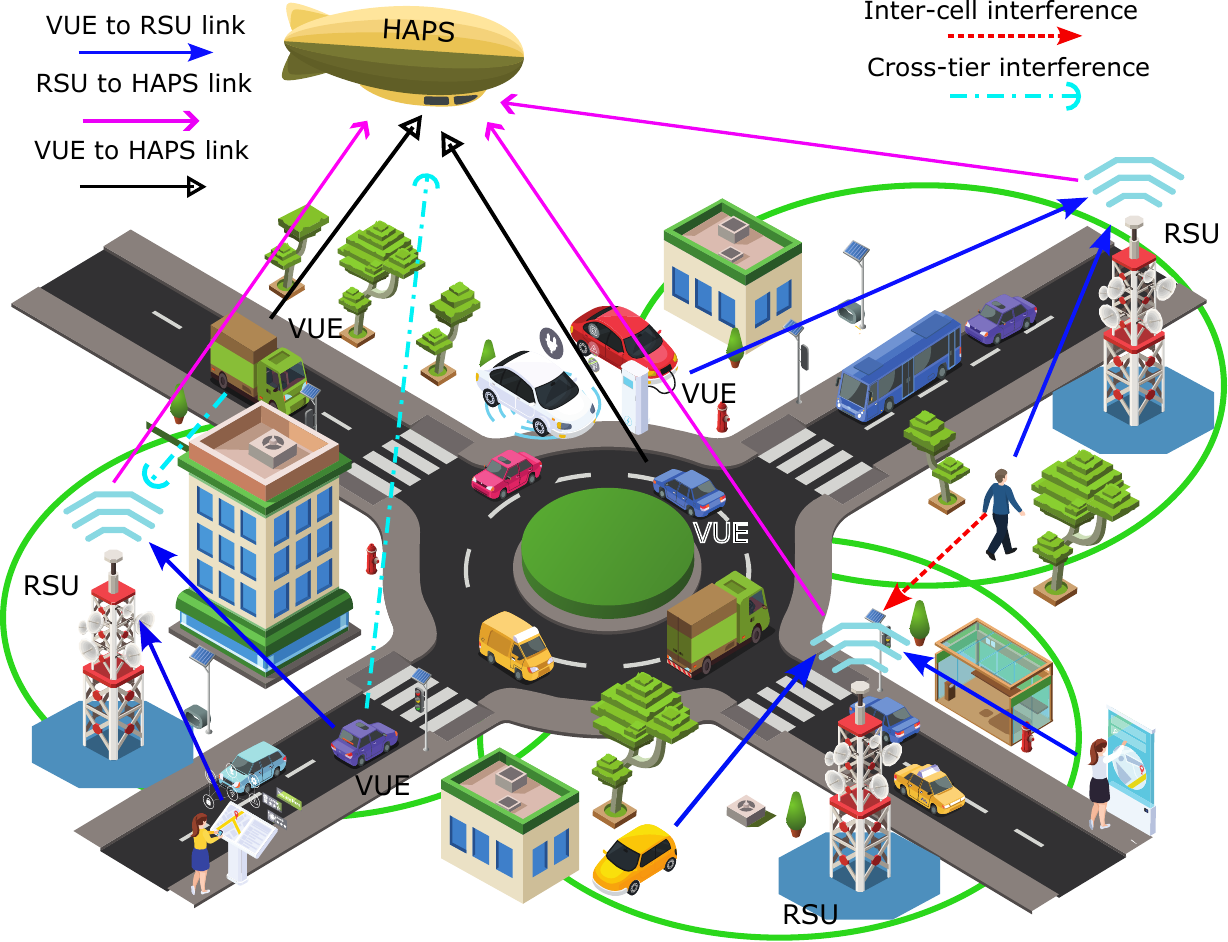}
	\caption{System Model}
	\label{fig:SM}
\end{figure*}
\subsection{Network Model}
As shown in Figure \ref{fig:SM}, this study focuses on the up-link communication scenario in HetNet, which consists of $M$ roadside units (RSUs) and one high altitude platform (HAP) denoted by $l$, serving $N$ vehicular user equipment (VUEs). RSUs play a crucial role in the network by providing a communication infrastructure for vehicles, enabling efficient communication between vehicles and the HAP. Without RSUs, the vehicles would have to rely solely on their own communication capabilities, leading to potential communication breakdowns and degraded network performance. The total system bandwidth, $B$, is divided into front-haul and back-haul links, with $(1-\eta)B$ allocated to front-haul communication between vehicles and RSUs, and $\eta B$ allocated to back-haul communication between RSUs and the HAP. The system incorporates wireless caching technology for data offloading in RSUs to meet network requirements. The NOMA technology enables multi-user transmission in this system, while communication between RSU-HAP links occurs over the C-band. The system model also considers several types of co-channel interference, including cross-tier HAPs interference, inter-cell interference, and intra-cell interference caused by NOMA. A comprehensive analysis of the system model is presented in subsequent sections, highlighting the importance of the proposed system design for efficient utilization of network resources and improved network performance.
\subsubsection{Terrestrial Communication Model} In a terrestrial communication network, NOMA technology allows an RSU to serve multiple VUEs \cite{khan2021energy}. The set of VUEs served by a particular RSU, denoted as $m$, determines the signal received by these VUEs at their associated RSU. Mathematically, this can be represented as:
\begin{equation}
\label{E1}
\begin{aligned}
y_{n, m}= &  \sqrt{p_{n, m}}h_{n, m} s_{n, m}+\!\widehat{\Phi_{n,m}^{I_1}}+\!\widehat{\Phi_{n,m}^{I_2}}+\widehat{\Phi_{n,m}^{I_3}}+\varsigma_n
\end{aligned}
\end{equation}
Following that, the equation \eqref{E1} describes the impact of different types of co-channel interference, including intra and inter-cell interference as well as cross-tier interference denoted by $\widehat{\Phi_{n,m}^{I_1}}, \widehat{\Phi_{n,m}^{I_2}},$ and $\widehat{\Phi_{n,m}^{I_3}}$, respectively. These sources of interference are mathematically represented as follows:
\begin{subequations}
    \begin{align}
        &\widehat{{\Phi_{n,m}^{I_1}}}=\sum_{{n'}\ne n \in L_m} \sqrt{p_{n',m}}h_{n,m}s_{n',m}.\\
        &\widehat{\Phi_{n,m}^{I_2}}=\sum_{j\in \mathbb{M}/m}\sum_{i\in L_j}\sqrt{p_{i,j}}h_{i,m}s_{i,j}.\\
        &\widehat{\Phi_{n,m}^{I_3}}=\sum_{i\in L_l} \sqrt{p_{i,l}}h_{i,m}s_{i,l}.
    \end{align}
\end{subequations}
In a NOMA-enabled HetNet, the transmission power from the $n^{th}$ VUEs towards the $m^{th}$ RSU and the HAPs is represented as $p_{n,m}$ and $p_{n,l}$, respectively. The channel coefficients between the $n^{th}$ VUE and $m^{th}$ RSU are denoted as $h_{n,m}$, while the signal transmitted from the $n^{th}$ VUE to the $m^{th}$ RSU and from the $j^{th}$ VUE to the HAPs are represented by $s_{n,m}$ and $s_{j,l}$, respectively. The channel between VUE-RSU/HAPs is modelled as additive white Gaussian noise (AWGN) and assumed to follow a complex Gaussian distribution, $\varsigma \sim \operatorname{CN}(0, \sigma^2)$, where $\sigma^2$ is the variance. Given these conditions, the signal-to-interference-plus-noise ratio (SINR) between the $n^{th}$ VUE and $m^{th}$ RSU can be calculated as follows:
\begin{equation}
\Gamma_{n, m}=\frac{\left|h_{n, m}\right|^2 p_{n, m}}{{\Phi_{n,m}^{I_1}}+\!{\Phi_{n,m}^{I_2}}+{\Phi_{n,m}^{I_3}}+\sigma^2},
\end{equation}
Where the intra, inter-cell, as well as cross-tier interference, are represented by
\begin{subequations}
    \begin{align}
        &\Phi_{n,m}^{I_1}=\sum_{{n'}\ne n \in L_m} {p_{n',m}}\left|h_{n,m}\right|^2.\\
        &\Phi_{n,m}^{I_2}=\sum_{j\in \mathbb{M}/m}\sum_{i\in L_j}{p_{i,j}}\left|h_{i,m}\right|^2\\
        &{\Phi_{n,m}^{I_3}}=\sum_{i\in L_l} {p_{i,l}}\left|h_{i,m}\right|^2.
    \end{align}
\end{subequations}
respectively. 
\par
Following that, the allocation coefficient, ${\eta}$, determines the portion of bandwidth allocated to the backhaul network. The user association between VUEs, RSUs, and HAP is denoted by the matrix $\bf{U}\left[u_{\mathrm{n}, m}\right]_{N \times M+l}$. This binary matrix indicates the connection between VUEs and RSU/HAPs, where $u_{n,m}=1$ if VUE $n$ is associated with RSU/HAPs, and 0 otherwise. By utilizing the user association information and the allocation coefficient, the transmission rate from VUE $n$ to RSU $m$ can be calculated based on the available bandwidth, channel quality, and other relevant factors. This rate represents the amount of data that can be transmitted over a certain period.
\begin{equation}
R_{n,m}=(1-{\eta}) B \log _2\left(1+\Gamma_{n, m}\right),
\end{equation}
Meanwhile, the overall achievable rate at the $m^{th}$ RSU can be expressed as follows:
\begin{equation}
R_m=\sum_{n \in \mathbb{N}} u_{n, m} R_{n, m} .
\end{equation}
\subsubsection{VUE-HAPs Communication} The signal received at the HAPs from the $n^{th}$ VUE, denoted by $y_{n,l}$, can be expressed mathematically as:
\begin{equation}
\begin{aligned}
y_{n,l}= &  \sqrt{p_{n,l} }h_{n,l} s_{n,l}+ \!\widehat{\Phi_{n,l}^{I_1}}+\!\widehat{\Phi_{n,l}^{I_2}}+\varsigma_l
\end{aligned}
\end{equation}
In this expression, $\widehat{\Phi_{n,l}^{I_1}}$ and $\widehat{\Phi_{n,l}^{I_2}}$ represent the intra-cell interference and the interference from other RSUs at the HAP, respectively. Mathematically, can be expressed as:
\begin{subequations}
    \begin{align}
        &\widehat{\Phi_{n,l}^{I_1}}=\sum_{n' \in L_l /n} \sqrt{p_{n', l}}h_{n,l} s_{n',l}\\
        &\widehat{\Phi_{n,l}^{I_2}}=\sum_{j\in \mathbb{M}}\sum_{i\in L_j} \sqrt{p_{i, j}}h_{i,l}  s_{i,l}
    \end{align}
\end{subequations}
Similarly, the channel coefficient between a VUE and the HAP is denoted by $h_{u,l}$. The SINR that the HAP receives from the VUE can be expressed as a function of this coefficient as follows:
\begin{equation}
\Gamma_{n,l}=\frac{\left|h_{n,l}\right|^2 p_{n,l}}{{\Phi_{n,l}^{I_1}}+\!{\Phi_{n,l}^{I_2}}+\sigma^2}
\end{equation}
Where,
\begin{subequations}
    \begin{align}
        &\Phi_{n,l}^{I_1}=\sum_{n' \in L_l /n} {p_{n', l}}\left|h_{n,l}\right|^2\\
        &\Phi_{n,l}^{I_2}=\sum_{j\in \mathbb{M}}\sum_{i\in L_j} {p_{i, j}}\left|h_{i,l}\right|^2.
    \end{align}
\end{subequations}
respectively. 
\par
Similarly, VUEs associated with HAPs are subject to both intra-HAPs and RSU-originated interference, which can impact their SINR. Furthermore, since the transmission power of HAPs' VUEs is constant, the SINR at the HAPs is also affected by cross-tier interference from RSUs' VUEs.
\subsubsection{Caching Model} 
Local caching is a popular feature implemented in RSUs to reduce network traffic and alleviate backhaul pressure by storing frequently requested data at the network's edge. This allows VUEs to access content directly from the local storage of RSUs instead of relying on the limited capacity of backhaul links. The effectiveness of the caching strategy is influenced by the power allocation values. The caching index, denoted as $\mathbb{X}\left[x_{n,m}\right]{N \times M}$, reflects the success of caching UE $n$'s content at RSU $m$ during the caching phase, where $x_{n,m}=1$ indicates successful caching and $x_{n,m}=0$ indicates otherwise. It is worth noting that each VUE's content can only be cached at a single RSU, and all RSUs must have the same buffer capacity. The constraints enforced by this approach can be expressed as $\sum_{m} x_{n, m} \leq 1$ and $\sum_{i \in\{L_m\}} x_{i,m} \leq x_{\max }$. 
\subsubsection{Back-haul Link Capacity Model} In the proposed system, as illustrated in Figure \ref{fig:SM}, VUEs access content via HAPs, and front-haul communication is constrained by the back-haul communication link between the RSUs and the HAPs. Caching ($x_{n,m}=1$) allows BSs to serve cached content to VUEs, reducing the back-haul burden and link overhead. The Orthogonal Frequency Division Multiple Access (OFDMA) technique is used for communication between RSUs and HAPs, as reported in \cite{yang2018flexible}. The system also specifies the Signal-to-Interference-plus-Noise Ratio (SINR) of the back-haul link from RSUs $m$ to the HAPs.
\begin{equation}
\Gamma_{m,l}=\frac{p_{m, l}\left|h_{m, l}\right|^2}{ \sum_{m' \in \mathbb{M}\backslash m } p_{m', l}\left|h_{m,l}\right|^2+\sigma^2} .
\end{equation}
In addition, the rate equation for the backhaul communication link from the $m^{th}$ RSU to the HAPs can be expressed as:
\begin{equation}
R_{m,l}=\Psi_m \log _2\left(1+\Gamma_{m,l}\right),
\end{equation}
where $\Psi_m=\frac{\left|L_m\right|-\sum_{i \in L_m} x_{i, m}}{\left|L_m\right|} \frac{{\eta} B}{M}$ and  $\left|L_m\right|=\sum_{i \in \mathrm{c}} u_{i, l}$. The achievable back-haul transmission rate is
\begin{equation}
R_{\mathrm{l}}=\sum_{m \in \mathbb{M}} R_{m,l}.
\end{equation}
\subsubsection{Mechanism of SIC Decoding} 
{\color{black}The proposed system utilizes the NOMA technique to address the challenges of multiplexing multiple signals in the same frequency band \cite{khan2022noma}. The technique relies on the SIC approach to prioritize the decoding process based on the received signal strength, with the user receiving the strongest signal being decoded first. This method is employed for both uplink and downlink transmissions, with the user with the best channel conditions receiving the highest priority for decoding in an uplink NOMA scenario with equal transmitter capabilities \cite{khan2021energy}. This approach enables the system to allocate resources more efficiently, leading to improved communication performance. In the proposed system model, various factors such as the inter-cell interference $\Phi_{n,m}^{I_1}$, interference from outside the cell $\Phi_{n,m}^{I_2}$, and interference from the HAPs to RSU $\Phi_{n,m}^{I_3}$ are considered in determining the channel state. The system's approach is formalized in a theorem presented in the paper as follows.}

\textbf{Theorem I}: For NOMA networks featuring VUE-RSU uplink and multiple cells, it is necessary for the channel coefficients of users $n_1$ and $n_2$ in cell $m$ (where $m \in \mathbb{M}$) to meet a specific requirement in order for the SIC technique to be applied successfully and eliminate $n_1$'s signal from $n_2$'s signal.
\begin{equation}
\left|h_{{n_1}, l}\right|^2 \geq\left|h_{{n_2}, m}\right|^2 .
\end{equation}
\textbf{Proof:} The power received by RSU $m$ from VUES $n_1$ and $n_2$ are as follows:
\begin{equation}
\begin{aligned}
& {n_1^r}=\left|h_{{n_1}, m}\right|^2\left(p_{{n_1}, l}+p_{{n_2}, m}\right)+\Phi_{n_1,m}^{I_2}+\Phi_{n_1,m}^{I_3}+\sigma^2 \\
& n_2^r=\left|h_{n_2, m}\right|^2\left(p_{n_2, m}+p_{n_2, m}\right)+\Phi_{n_2,m}^{I_2}+\Phi_{n_2,m}^{I_3}+\sigma^2
\end{aligned}
\end{equation}
The condition for successful implementation of the SIC technique in uplink RSU-HAPs multiple-cell NOMA networks, used to decode and eliminate user $n_1$'s signal from user $n_2$'s signal, can be expressed as follows: $\Phi_{n_1,m}^{I_2}=\Phi_{n_2,m}^{I_2}=\sum_{j\in \mathbb{M}/m}\sum_{i\in L_j}{p_{i,j}}\left|h_{i,m}\right|^2$ and $\Phi_{n_1,m}^{I_3}=\Phi_{n_2,m}^{I_3}==\sum_{i\in L_l} {p_{i,l}}\left|h_{i,m}\right|^2$. Analysis shows that for user $n_1$'s signal to be decoded successfully, its received signal must be greater than or equal to that of user $n_2$, i.e., $\left|h_{n_1, m}\right|^2 \geq\left|h_{n_2, m}\right|^2$. This completes the proof. Arranging the channel coefficients of the VUEs connected to the RSU $m$ in ascending order facilitates the systematic implementation of the SIC technique. By decoding the signal from the VUEs with the superior channel condition first, communication performance can be improved.
\begin{equation}
W\left(L_m\right) \triangleq\left|h_{1, m}\right|^2 \geq\left|h_{2, m}\right|^2 \geq \cdots \geq\left|h_{N, k}\right|^2, \forall m .
\end{equation}
Therefore, based on the above assumption, the SINR of RSU from VUEs $n$ can be expressed as $J_n$, where $J_n$ represents the set $\{1,2, \cdots, n\}$.

\begin{equation}
\Gamma_{n,m}=\frac{\left|h_{n,m}\right|^2 p_{n,m}}{ \sum_{i \in L_m \backslash J_n} p_{i, m}+\Phi_{n,m}^{I_2}+\Phi_{n,m}^{I_3}+\sigma^2}
\end{equation}

\subsection{Problem Formulation}
In a NOMA-enabled vehicular-aided HetNet, the primary objective is to achieve high transmission rates while limiting the impact of cross-tier interference on the QoS requirements of VUEs connected to a HAP. NOMA technology enables multiple users to share the same frequency and time resources using power-domain multiplexing, but cross-tier interference can still occur and affect the QoS of VUEs. To address this challenge, the utility function for VUEs must consider both the achievable transmission rate and the impact of cross-tier interference while taking into account the specific QoS requirements of the VUEs and the methods used to calculate the utility function. Therefore to address this, We formulate a weighted sum of the achievable transmission rate and a penalty function for the interference level. By adjusting the weight factors, the system can balance the two objectives based on the system's priorities. This balance is critical for providing reliable and efficient communication services in vehicular networks, as demonstrated mathematically as follows.
\begin{equation}
\label{E18}
F_{n,m}=\sum_{n \in L_m}\left(R_{n,m}-\Omega\left|h_{n,l}\right|^2 p_{n,m}\right),
\end{equation}
{\color{black}The interference pricing factor $\Omega$ plays a crucial role in balancing the trade-off between transmission rate and cross-tier interference. While a higher transmission rate increases the utility, cross-tier interference can significantly reduce it. In our simulations, we adopt a dynamic approach for setting the parameter $\Omega$ to adapt to real-time network conditions and user requirements. Therefore, the total utility of the system is the sum of the utilities of each RSU, represented by $F_m$.}
\begin{equation}
F_m= \sum_{n \in L_m} F_{n,m}.
\end{equation}
The problem of resource allocation in the uplink NOMA-enabled vehicular-aided HetNet can be mathematically formulated as an optimization problem. The objective is to maximize the system utility, subject to constraints such as power budget, QoS requirements of users, and interference limits. Specifically, let $F_m$ denote the utility of the RSU, the problem can be expressed as:
\begin{subequations}
\label{OP1}
\begin{align}
& \max_{\{{\textbf{U}}, {\eta}, \textbf{{P}}\}} \sum_{n \in L_m}\left(R_{n,m}-\Omega_t\left|h_{n,l}\right|^2 p_{n,m}\right) \label{OF1} \\
& C 1: R_{n,m}>R_n, \quad \forall m \in \mathbb{M}, \forall n \in \mathbb{N}, \label{C5}\\
& C 2: \sum_{n' \in L_m}\left(1-g_{n', m}\right) R_{n', m}<R_{m, l}, \forall m \in \mathbb{M} \label{C7}.\\
& C 3: W\left(L_m\right), \forall m \in \mathbb{M}+l, \label{C1}\\
& C 4: u_{n,m} \in\{0,1\}, \forall m \in \mathbb{M}+1, \forall n \in \mathbb{N},\label{C2} \\
& C 5: {\sum}_{j=1}^{M+l} u_{n, j}=1, \quad \forall n \in \mathbb{N}, \label{C3} \\
& C 6: p_{n,m} \in\left[0, P_{\max }\right], \forall n \in \mathbb{N},\forall m \in \mathbb{M}+l, \label{C4}\\
& C 7: {\eta} \in(0,1), \label{C6}
\end{align}
\end{subequations}
The resource allocation problem in the uplink NOMA-enabled vehicular-aided HetNet network is a complex optimization task that aims to jointly optimize the allocation of resources, including the VUE-RSU/HAPs association variable $\textbf{U}$, the indicator for bandwidth assignment ${\eta}$, and the transmission power vector $\textbf{P}$. The problem is subject to a set of constraints that must be satisfied to ensure optimal resource allocation.
\begin{enumerate}
    \item Constraint \eqref{C5} is imposed to guarantee that each VUEs quality of service (QoS) requirement is met.
    \item Constraint \eqref{C7} is imposed to ensure that the total rate achieved by each cell $m$ does not exceed its available backhaul link rate.
    \item Constraint \eqref{C1} ensures that the decoding order of users in each cell $m$ is maintained through the use of the access point-user association matrix $\mathrm{U}\left[u_{n,m}\right]{N \times l}$.
    \item Constraint \eqref{C2} restricts each VUEs only to be associated with one access point, either an RSU $m$ or HAPs $l$, with $u{n, m}=1$ indicating association and $u_{n,m}=0$ indicating no association.
    \item Constraint \eqref{C3} enforces that each VUE can only be served by one AP at a time.
    \item The power constraint is defined in \eqref{C4}.
    \item Constraint \eqref{C6} represented the bounds for the back-haul bandwidth allocation factor.
\end{enumerate}
These constraints ensure that the  optimal allocation of resources in the up-link NOMA-enabled vehicular-aided HetNet network is optimized and meets the necessary requirements.
\par
The optimization problem described in equation \eqref{OP1} presents a challenging mixed-integer nonlinear programming problem that is difficult to solve and optimize. The problem is characterized by non-convexity and NP-hard complexity, making it challenging to obtain a globally optimal solution. Additionally, the user association strategy dynamically affects the channel conditions, further complicating the optimization problem. To overcome these challenges, the joint optimization problem is decomposed into several distinct subproblems, resulting in improved solution efficiency. This approach enables the optimization of each subproblem separately, leading to better convergence and reducing the computational complexity of the overall problem. By breaking down the joint optimization problem, it becomes possible to solve for the optimal resource allocation with reduced computational resources and time.

\section{ Framework for Optimal Allocation of resources}
The optimization problem specified in equation \eqref{OP1} is a challenging mixed-integer nonlinear programming issue with computational difficulties and difficulties in optimizing the solution. Due to non-convexity and NP-hard complexity, obtaining a globally optimal solution is challenging. Additionally, the user association strategy dynamically influences channel conditions, adding further complexities. To address these issues, the joint optimization problem is divided into three sub-problems to improve solution efficiency. The first sub-problem develops a methodology for UE association based on the current caching state and user preferences with fixed power-to-bandwidth ratios. The second sub-problem focuses on bandwidth allocation. Finally, the third sub-problem deals with power allocation. An iterative algorithm is proposed to find the optimal power allocation, using information from previous sub-problems on UE association and allocation of bandwidth. The methodology for each sub-problem is detailed in the following section.
\subsection{User Association Method}
The sub-problem for UE association can be expressed as follows: 
\begin{subequations}
\label{OP2}
\begin{align}
& \max _{\{{\textbf{U}}\}} \sum_{n \in L_m}\left(R_{n,m}-\Omega\left|h_{n,l}\right|^2 p_{n,m}\right) \label{OF2} \\
& C 1: W\left(L_m\right), \forall m \in \mathbb{M}+l, \label{C1'}\\
& C 2: u_{n,m} \in\{0,1\}, \forall m \in \mathbb{M}+1, \forall n \in \mathbb{N},\label{C2'} \\
& C 3: {\sum}_{j=1}^{M+l} u_{n, j}=1, \quad \forall n \in \mathbb{N}, \label{C3'} 
\end{align}
\end{subequations}
To address the UE association optimization sub-problem stated in equation \eqref{OP2}, an algorithm is utilized that considers caching, preference relations, and swapping. Notably, the algorithm doesn't account for the QoS requirements of the UE and the back-haul link, which provides a broader user swapping range and facilitates obtaining globally optimal solutions.

\subsubsection{Preparation process} {\color{black} In the proposed resource allocation method for the NOMA-enabled vehicular-aided HetNet network, the selection of VUEs connected to the HAPS is based on the ratio of channel coefficients, $\kappa_n=\frac{\left|h_{n,l}\right|^2}{\left|h_{n,m}\right|^2}$. This ratio is computed for each VUE and sorted in descending order. The top $L_l$ VUEs with the highest $\kappa_n$ values are then designated as the HAP users, while the remaining VUEs connect to the RSUs. The selection of the RSU is determined by the VUEs' preference list, which is sorted according to the channel coefficients, with the position of RSU $m$ in the preference list indicated by $l_{m,n}$. The proposed method provides a systematic approach to allocate resources in vehicular-aided HetNet networks while considering VUE preferences and channel conditions. This approach can help optimize the performance of the network while ensuring fair allocation of resources among the VUEs.
\subsubsection{Sense and  Action} The VUEs connected to the RSU transmit request signals in accordance with their preference list. The decision rule for any $m_1, m_2 \in M$ where $m_1 \neq m_2$ is stated as follows:}
\begin{equation}
\label{E19}
\begin{aligned}
L_{m_1,n} & \succ L_{m_2} \\
& \Leftrightarrow N_{n,m_2}^B\left(\left\{L_{m_1}, L_{m_2}\right\}\right) \\
& <N_{n, {m_1}}\left(\left\{L_{k_1} \cup\{n\}, L_{{m-2}} \backslash\{n\}\right\}\right) .
\end{aligned}
\end{equation}
Equation \eqref{E19} presents the preference of VUE $n$ between RSU $m_1$ and $m_2$, which is based on the RSU that can offer higher utility. To make the best use of caching resources, VUEs prefer to connect to RSUs that have their desired content cached. However, the long communication distance can lead to a high path loss and reduce the benefits of caching. To address this, the utility function is weighted by a factor $\alpha$. Thus, when $x_{n,m}=1$, the evaluation function can be expressed as:
\begin{subequations}
\begin{align}
\omega_{n,v}&=\frac{x_{n,m}\left(1+\Gamma_{n,m}\right)^{\alpha(1-\eta)}}{\Xi_v},\forall v \in \mathbb{V}\\
\Xi_v&=\left(\frac{\left(1+\Gamma_{n,v}\right)^{\alpha(1-\eta)}}{\left(1+\Gamma_{v, l}\right)^{\frac{(1-\alpha) \|}{L_z / m}}}\right)
\end{align}
\end{subequations}
The proposed method in this study involves assigning a weighting factor, denoted by $\alpha$, to the front-haul link. The set $\mathbb{V}$ consists of all RSUs that are closer to VUE $n$ than RSU $m$. The decision of VUE $n$ to connect to a particular RSU depends on the value of $\omega_{n,v}$. If the value of $\omega_{n,v}$ is greater than or equal to 1 for all RSUs in $\mathbb{V}$, it implies that the revenue generated by caching compensates for the increased path loss due to long-distance communication, and VUE $n$ connects to RSU $v$. On the other hand, if the value of $\omega_{n,v}$ lies between 0 and 1 for some RSUs in $\mathbb{V}$, VUE $n$ selects the RSU with the lowest $\omega_{n,v}$ as it generates more revenue than RSU $m$. The action of VUE $n$ is denoted by $\Lambda_n$, and the above conditions can be expressed as follows:
\begin{equation}
\label{E21}
\Lambda_n= \begin{cases}u_{n,m}=1, & \text{Case 1}, \\ 
\sum_{v \in \mathbb{V}} u_{n,v}=1, & \text{Case 2}, \\ 
\text { judge and swap by (19), } & \text{Case 3}.\end{cases}
\end{equation}
Similarly, the cased can be represented as follows\\
\begin{equation}
\begin{aligned}
    \text {Case 1:  if }& x_{n,m}=1,  \omega_{n,v} \geq 1, \forall v \in \mathbb{V}, \forall m \in \mathbb{M}\\
    \text { Case 2: if }& x_{n,m}=1,  0<\omega_{n,v}<1, \exists v \in \mathbb{V}, \forall m \in \mathbb{M}\\
    \text { Case 3: if  }& x_{n,m}=0, \forall m \in \mathbb{M} 
\end{aligned}
\end{equation}

\subsubsection{Swap Matching process} For any pair of RSU $m_1, m_2 \in \mathbb{M}$, where $m_1 \neq m_2$, and any pair of VUEs $n_1, n_2 \in \mathrm{N}$, where $n_1 \neq n_2$ such that $u_{n_1, m_1}=1$ and $u_{n_2, m_2}=1$, the swapping matching process is explained as follows..
\begin{equation}
\label{E22}
\begin{aligned}
\{L\}_{n_1}^{n_2} & =\{L\} \backslash\left\{L_{m_1}, L_{m_2}\right\} \\
& \cup\left\{L_{m_1} \backslash\{n_1\} \cup\{n_2\}\right\} \\
& \cup\left\{L_{m_2} \backslash\{n_2\} \cup\{n_1\}\right\} .
\end{aligned}
\end{equation}
Similarly, the rules for the swapping can be expressed as follows:
\begin{equation}
\label{E27}
\begin{aligned}
\{L\}_{\mathrm{u} 1(n_1, n_2)}^{u, 2} & \succ\{L\} \\
& \Leftrightarrow \sum_{m \in \mathbb{M}} F_m(\{L\})<\sum_{k \in \mathbb{M}}F_m\left(\{L\}_{n_1}^{n_2}\right) .
\end{aligned}
\end{equation}

In Equation \ref{E27}, the system utility function $F_m$ is defined as the sum of the revenue generated by all VUEs connected to RSU $m$, subtracted by the product of the channel coefficient and transmit power of each VUE. The matching ${L}$ is only updated to ${L}_{n_1}^{n_2}$ if the new matching results in a higher system utility compared to the previous matching.
\subsubsection{End of the Algorithm}
The algorithm consists of two processes: the sense and action process and the swap matching process. The sense and action process involves continuously optimizing the VUE association matrix by calculating the user utility using Equation \eqref{E19} and making changes until no VUE wants to switch access points. In the swap matching process, two VUEs are randomly selected, and the utility function is evaluated using Equation \eqref{E22} to find the optimal AP-UE association. This process continues until convergence is reached.

\subsection{Bandwidth Assignment}
Decoupling the original optimization problem \eqref{OP1} into a sub-problem for bandwidth allocation yields the following mathematical expression:
\begin{subequations}
\label{OP3}
\begin{align}
& \max _{\{\eta\}} \sum_{m \in \mathbb{M}} \sum_{n \in L_m}\left(R_{n,m}-\Omega\left|h_{n,l}\right|^2 p_{n,m}\right) \\
& C 1: \sum_{n' \in L_m}\left(1-g_{n', m}\right) R_{n', m}<R_{m, l}, \forall m \in \mathbb{M} \label{C111}.\\
& C 2: W\left(L_m^*\right), \forall m \in \mathbb{M}+l, \label{C222}\\
& C 3: {\eta} \in(0,1), \label{C333}
\end{align}
\end{subequations}
The term $W\left(L_m^*\right)$ in the equation refers to the decoding sequence of cell $m$ with the optimal VUE association $\mathbb{U}^{[u_{n,m}^*]}$. Using this, the closed-form solution for optimizing the bandwidth as given in Equation \eqref{OP3} can be obtained. The solution involves maximizing the system throughput by allocating bandwidth to each cell based on the optimal VUE association. The optimization problem is decoupled into a sub-problem for bandwidth allocation, and the solution is obtained by iteratively adjusting the bandwidth allocations until the optimal solution is achieved. The mathematical details of this solution are provided below.

\textbf{Definition 1}: In the optimization sub-problem \eqref{OP3} for bandwidth allocation, the optimal value of $\eta$ can be found by maximizing $JB_m$ over all $m \in \mathbb{M}$. The expression for $JB_m$ is given as follows:
\begin{equation}
J B_m\!=\!\frac{\sum_{i \in L_m^*} x_{i,m}'\log _2\left(1+\Gamma_{i,m}\right)}{\Psi_m\Theta_m \!+\!\sum_{i \in L_m^*}\!\! x_{i, m}'\log _2\left(1\!+\!\Gamma_{i,m}\right)},\! \forall m \!\in \mathbb{M} .
\end{equation}
Where, $\Psi_m=\frac{\left|L_m^*\right|-\sum_{i \in L_m^*} x_{i,m}}{\left|L_m^*\right|} \frac{1}{M}$, 
$x_{i,m}'=1-x_{i, m}$ and $\Theta_m=\log _2\left(1+\Gamma_{m}^o\right)$.  Similarly, 
\begin{equation}
\small
\label{E30}
\begin{aligned}
& \sum_{i \in L_m^*}{x_{i,m}'}(1\!-\!\eta) \!B\! \log _2\left(1\!+\!\Gamma_{i,m}\right) \!\!\leq\!\! \Psi_m \eta{B} \log _2\left(1\!+\!\Gamma_{k, s}\right) \\
&\!\! \Rightarrow\! \eta \!\geq \frac{\sum_{i \in L_m^*}{x_{i,m}'} \log _2\left(1+\Gamma_{i,m}\right)}{\!\Psi_m\! \log _2\left(1+\Gamma_{m,l}\right)+\sum_{i \in L_m^*}{x_{i,m}'} \log _2\left(1+\Gamma_{i,m}\right)}\\
&=J B_m \Rightarrow \eta \geq \max _m J B_m .
\end{aligned}
\end{equation}
\textbf{Proof}: In the optimization subproblem represented by equation \eqref{OP3}, the optimal value of the allocation parameter $\eta$ can be obtained as the maximum of $JB_m$ for all $m \in \mathbb{M}$. For each cell $m \in \mathbb{M}$, the value of $JB_m$ is defined based on the conditions \eqref{C222} and \eqref{C333}. Using condition \eqref{C333}, the expression in equation \eqref{E30} can be derived for each $m \in \mathbb{M}$. Consequently, the optimization subproblem \eqref{OP3} can be transformed and formulated based on the derived expression in equation \eqref{E30} as follows.
\begin{equation}
\label{E31}
\begin{aligned}
& \max _{\{\eta\}} \sum_{m\in \mathbb{M}} \sum_{n \in L_m}\left(R_{n,m}-\Omega\left|h_{n,l}\right|^2 p_{n,m}\right) \\
& \text { s.t. } \max _m J B_m \leq \eta<1, \quad \forall n \in \mathbb{N}, m \in \mathbb{M} .
\end{aligned}
\end{equation}
The solution to the bandwidth allocation optimization subproblem, represented by equation \eqref{E31}, can be obtained by computing the optimal value of $\eta^*$ that corresponds to the lower bound of the monotonically decreasing utility function. This can be achieved using bisection search or other efficient numerical optimization techniques. Once the optimal value of $\eta^*$ is obtained, the optimal bandwidth allocation can be obtained using equation \eqref{E30}.
\subsection{Power Allocation}
The mathematical expression for the sub-problem to find the transmission power is as follows:
\begin{subequations}
\label{OP5}
\begin{align}
& \max _{\{\textbf{P}\}} \sum_{m \in \mathbb{M}} \sum_{n \in L_m}\left(R_{n,m}-\Omega\left|h_{n,l}\right|^2 p_{n,m}\right) \\
& C 1: R_{n,m}>R_n, \quad \forall m \in \mathbb{M}, \forall n \in \mathbb{N}, \label{C1a}\\
& C 2: W\left(L_m^*\right), \forall m \in \mathbb{M}+l, \label{C2a},\\
& C 3: p_{n,m} \in\left[0, P_{\max }\right], \forall n \in \mathbb{N},\forall m \in \mathbb{M}+l. \label{C3a}
\end{align}
\end{subequations}
The objective function of the optimization problem is non-convex and can be represented as $\sum_{m \in \mathbb{M}} \sum_{n \in L_m}\left(R_{n,m}-\omega\left|h_{n,l}\right|^2 p_{n,m}\right)$ with respect to $p_{n,m}$. To address this issue, the objective function is reformulated as follows:
\begin{equation}
\label{E33}
\begin{aligned}
\max _{\{\textbf{P}\}} & \left\{\left(1-\eta^*\right) B \sum_{m \in \mathbb{M}} \sum_{n \in L_m^*} \log _2\left(\Upsilon_{n,m}^1\right)\right. \\
& \left.-\left(1-\eta^*\right) B \sum_{m \in \mathbb{M}} \sum_{n \in L_m^*} \log _2\left(\Upsilon_{n,m}^2\right)\right.\\
&\left.-\omega \sum_{n \in \mathbb{M}} \sum_{n \in L_m^*}\left|h_{n,l}\right|^2 p_{n,m}\right\} .
\end{aligned}
\end{equation}
Where, $\Upsilon_{n,m}^1=\left|h_{n,m}\right|^2 p_{n,m}+\left|h_{n,m}^B\right|^2 \sum_{i \in L_m^* \backslash J_n} p_{i,m}+\Phi_{n,m}^{I_2}+\Phi_{n,m}^{I_3}+\sigma^2$
$=\left|h_{n,m}^B\right|^2 \sum_{i \in L_m^* \backslash J_n} p_{i,m}+\Phi_{n,m}^{I_2}+\Phi_{n,m}^{I_3}+\sigma^2$
Subsequently, the second term of the objective function, denoted as \eqref{E33}, exhibits non-convexity and requires transformation into a convex form. This non-convex problem can be effectively addressed using the successive convex approximation approach, which has been shown to converge well and adhere to the Karush-Kuhn-Tucker (KKT) condition \cite{zhang2020energy,lai2016joint}. The inequality mentioned above can be approximated to the upper bound of the logarithmic function \cite{boyd2004convex}, which converges to $t^i=t_0^i$.
\begin{equation}
\small
\label{E34}
\begin{aligned}
-\gamma= & -\sum_{m \in \mathbb{M}} \sum_{n \in L_m^*} \log _2\left(\left|h_{n,m}\right|^2 \sum_{i \in L_m \backslash J_n} p_{i,m}+\Phi_{n,m}^{I_2}\right. \\
& \left.++\Phi_{n,m}^{I_3}+\sigma^2\right) .
\end{aligned}
\end{equation}
We define it as follows.\\
\textbf{Definition 2}: Lower bounds at $-\bar{\gamma}$ can be obtained via the non-convex term $-\gamma$ of \eqref{E34}, which converges at the local point$p_{n,m}=$ $p_{n,m}[t]$.
\begin{equation}
\small
\label{E35}
\begin{aligned}  
\tilde{\gamma}= & \sum_{n \in L_m^*} \log _2\left({\Phi_{n,m}^{I_1}}[t]+\!{\Phi_{n,m}^{I_2}}+{\Phi_{n,m}^{I_3}}+\sigma^2\right) \\
& +\sum_{j \in \mathbb{M} \backslash\{m\}} \sum_{i \in L_j^*} \log _2\left({\Phi_{j,i}^{I_1}}+\!{\Phi_{j,i}^{I_2}}[t]+{\Phi_{j,i}^{I_3}}+\sigma^2\right) \\
& +\frac{1}{\ln 2} \sum_{n \in L} \sum_{i \in J_{n-1}} \frac{\left|h_{i, m}\right|^2\left(p_{n,m}-p_{n,m}[t]\right)}{{\Phi_{i,m}^{I_1}}[t]+\!{\Phi_{i,m}^{I_2}}+{\Phi_{i,m}^{I_3}}+\sigma^2} \\
& +\frac{1}{\ln 2} \sum_{n \in L_k^*} \sum_{j \in \backslash\{m\}} \sum_{i \in M_j^*} \frac{\left.\left|h_{n,j}\right|^2\left(p_{n,m}-p_{n,m}[t]\right]\right)}{\!{\Phi_{n,m}^{I_2}}[t]+{\Phi_{n,m}^{I_3}}+\sigma^2} .
\end{aligned}
\end{equation}
As described by \eqref{E35}, the non-convex and non-linear optimization problem \eqref{OP5} transformed effectively  and expressed in the more trackable form \eqref{OP6} as indicated below. Subsequently, a method to find the optimal best solution for \eqref{OP6} is proposed through an iterative power allocation scheme.
\begin{subequations}
\label{OP6}
\begin{align}
 \max _{\{\textbf{P}\}} & \left\{\left(1-\eta^*\right) B \sum_{m \in \mathbb{M}} \sum_{n \in L_m^*} \log _2\left(\Upsilon_{n,m}^1\right)\right. \\
& \left.-\left(1-\eta^*\right) B \tilde{\gamma}\right.\left.-\omega \sum_{n \in \mathbb{M}} \sum_{n \in L_m^*}\left|h_{n,l}\right|^2 p_{n,m}\right\} .\\
& C 1: R_{n,m}>R_n, \quad \forall m \in \mathbb{M}, \forall n \in \mathbb{N}, \label{C1b}\\
& C 2: W\left(L_m^*\right), \forall m \in \mathbb{M}+l, \label{C2b},\\
& C 3: p_{n,m} \in\left[0, P_{\max }\right], \forall n \in \mathbb{N},\forall m \in \mathbb{M}+l. \label{C3b}
\end{align}
\end{subequations}
The optimization problem \eqref{OP6} is solved through an iterative power allocation scheme. In each iteration, an initial power value $p_{n,m}[l]$ is specified, whereas the solution for the transmission power is calculated using the standard optimization toolbox e.g. interior point method. The computed power at the current iteration is then treated as the initial power value for the next iteration, $p_{n,m}[t+1]$. The algorithm is executed repeatedly until convergence is achieved. The successive convex approximation approach used in this iterative power allocation scheme has been proven to provide a good convergence, satisfying the KKT conditions and yielding an effective solution for the non-convex optimization problem \eqref{OP6}.

\begin{algorithm2e}[t]
\SetAlgoLined
\textbf{input: } $N \leftarrow$ UEs, $M \leftarrow$ Base Station, $l \leftarrow$ satellite, $B \leftarrow$ Bandwidth \;
\textbf{Initialization: } $W(L_m) \leftarrow$ decoding order and UE priority list. \;
\textbf{Execution: }\;
\While{{Until Converge}}{
     $L_l \leftarrow$ Calculate $\{\kappa_n, 'Descend'\}$ / Satellite UEs. \;
    Update the UE association matrix $\textbf{U}$ \;
    \While{{Until Converge}}{
    \ForEach{$n \in \{1,\cdots, N'\}$}{
        \If{$x_{n,m}=1$}{
            Find$(\mathbb{V}) \rightarrow \text{claulate} \rightarrow \Omega$  \;
        }
        Update \textbf{U}$\leftarrow $ calculate \eqref{E21}\;
        Update $W(L_m)$ $\leftarrow $ by \textbf{U} calculated in \textbf{step 12}\;
        }}

         \While{{Until Converge}}{
         $[n_1,n_2] \leftarrow $\text{ Randomly selected users such that} $n_1^m \ne n_2^m$\;
         $\{M\}_{n_1}^{n_2}\leftarrow$ solve \eqref{E22}
         $\textbf{U} \leftarrow$ solve \eqref{E27}
    }
 }
\caption{\textbf{Framework for UE Association}}\label{algo:ALO1}
\end{algorithm2e}

\begin{algorithm2e}[t]
\small
\SetAlgoLined
\textbf{Initialization: } $\textbf{P}_o \leftarrow$ Transmission Power, $t_{max}$ , $F_o$, $\Omega \leftarrow \text{Initial Value}$ \;
\textbf{Execution: }\;
\While{{$t\le t_{max} \text{or} error \le \epsilon$}}{

    \ForEach{$n \in \{1,\cdots, N'\}$}{
        \If{$x_{n,m}=1$}{
        $p_{n,m} \leftarrow$ Calculate the power using \eqref{OP6} \;
        }
    }  
    Update Power $p_{n,m}[t+l]\leftarrow p_{n,m}[t]$, \;
    Select $\Omega$ dynamically based on real-time network conditions and user requirements\;

    \If{Network is congested}{
        $\Omega \leftarrow \Omega + \delta$ \;
    }\ElseIf{Network interference is high}{
        $\Omega \leftarrow \Omega - \delta$ \;
    }\Else{
        $\Omega$ remains unchanged\;
    }
    F[t] = solve the Utility Function with the current $\Omega$ \;
    error = F[t] - F[t-1] \;
 }
\caption{\textbf{Dynamic Transmission Power Allocation with $\Omega$ Selection}}\label{algo:ALO2}
\end{algorithm2e}

\section{Algorithm Design for Resource Allocation}
To solve the optimization problem, the proposed algorithm adopts a two-stage approach. The first stage optimizes the user association, bandwidth allocation, and transmission power, while the second stage further improves the solution obtained in the first stage by performing AP switching. The algorithm is designed to iteratively perform the two stages until convergence is achieved.
\par
The first stage of the algorithm is implemented using the alternating optimization technique, where each variable is optimized sequentially while keeping the others fixed. The user association and bandwidth allocation are optimized jointly, while the transmission power is optimized separately. The user association and bandwidth allocation optimization problem is solved using the subgradient method, while the transmission power optimization problem is solved using the successive convex approximation approach.
\par
The second stage of the algorithm involves randomly selecting a VUE and evaluating the potential utility gain of switching to another AP. If the utility gain is positive, the VUE is switched to the new AP, and the optimization problem is solved again to update the user association and transmission power. This process continues until no further AP switching results in a positive utility gain.
\subsection{Complexity Analysis}
The computational complexity of the proposed algorithm depends on the number of VUEs and APs, as well as the convergence criteria. The user association and bandwidth allocation optimization subproblem have a complexity of $O\left(N_{\mathrm{UE}}N_{\mathrm{AP}}\right)$, while the transmission power optimization subproblem has a complexity of $O\left(N_{\mathrm{UE}}N_{\mathrm{AP}}\log\frac{1}{\epsilon}\right)$, where $\epsilon$ is the accuracy of the solution. The complexity of the AP switching process is $O\left(N_{\mathrm{UE}}^2N_{\mathrm{AP}}\right)$. Therefore, the overall complexity of the proposed algorithm is $O\left(T\left(N_{\mathrm{UE}}N_{\mathrm{AP}} + N_{\mathrm{UE}}N_{\mathrm{AP}}\log\frac{1}{\epsilon} + N_{\mathrm{UE}}^2N_{\mathrm{AP}}\right)\right)$, where $T$ is the number of iterations required to achieve convergence.

\begin{figure}[!t]
	\centering
	\includegraphics[width=0.8\linewidth]{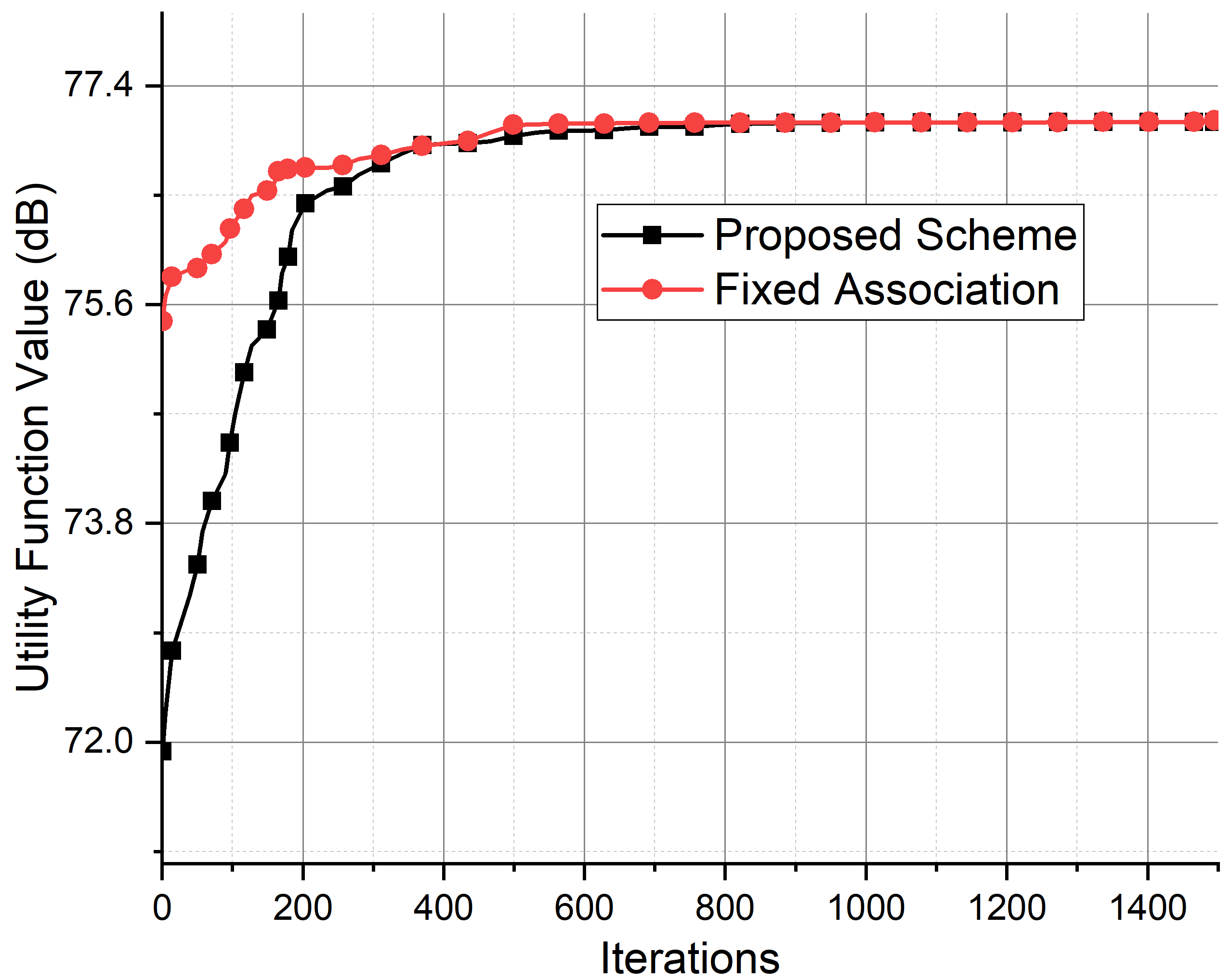}
	\caption{Convergence of Algorithm 1}
	\label{fig:R1}
\end{figure}
\begin{figure}[!t]
	\centering
	\includegraphics[width=0.8\linewidth]{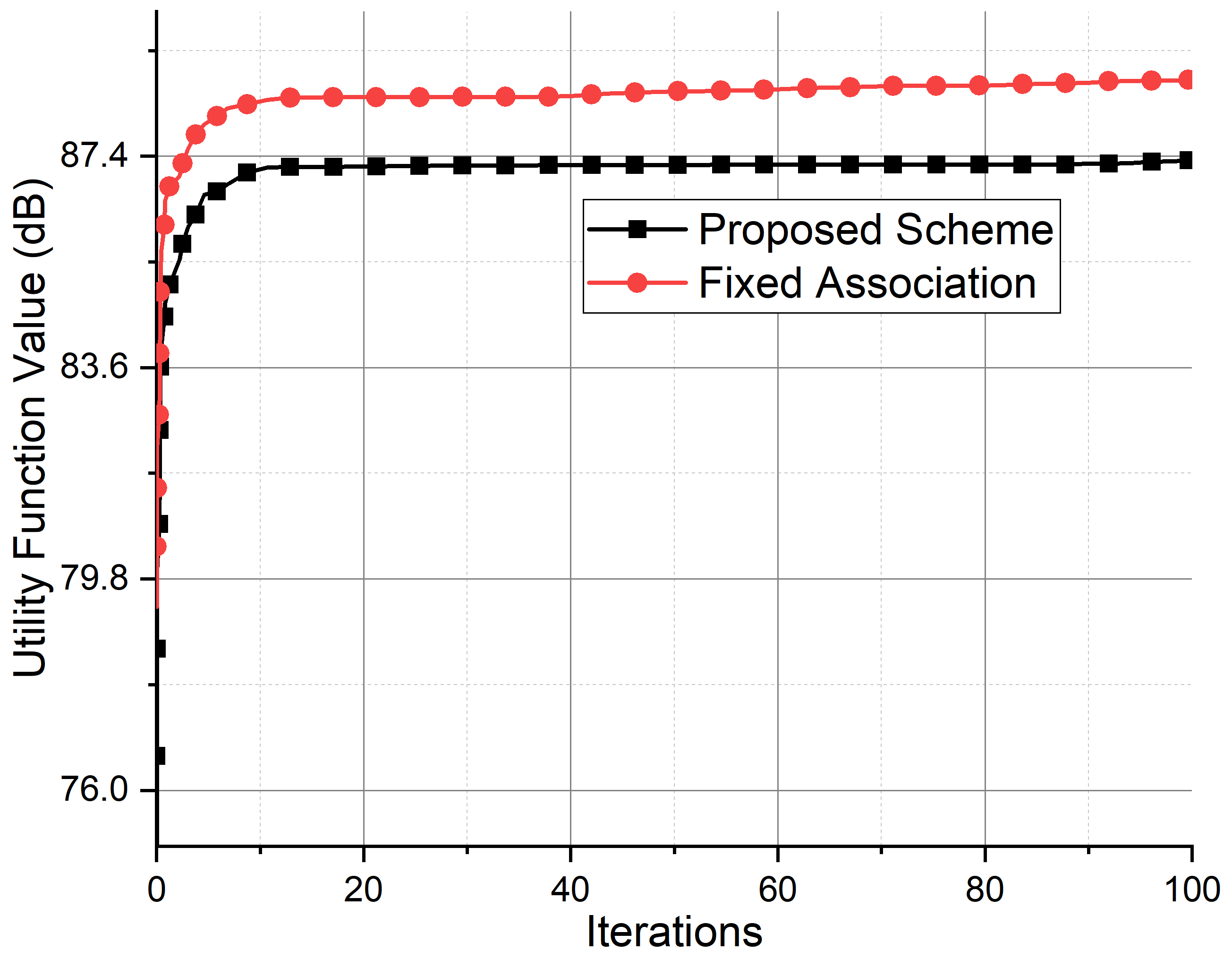}
	\caption{Convergence of Algorithm 2}
	\label{fig:R2}
\end{figure}
\begin{figure}[!t]
	\centering
	\includegraphics[width=0.9\linewidth]{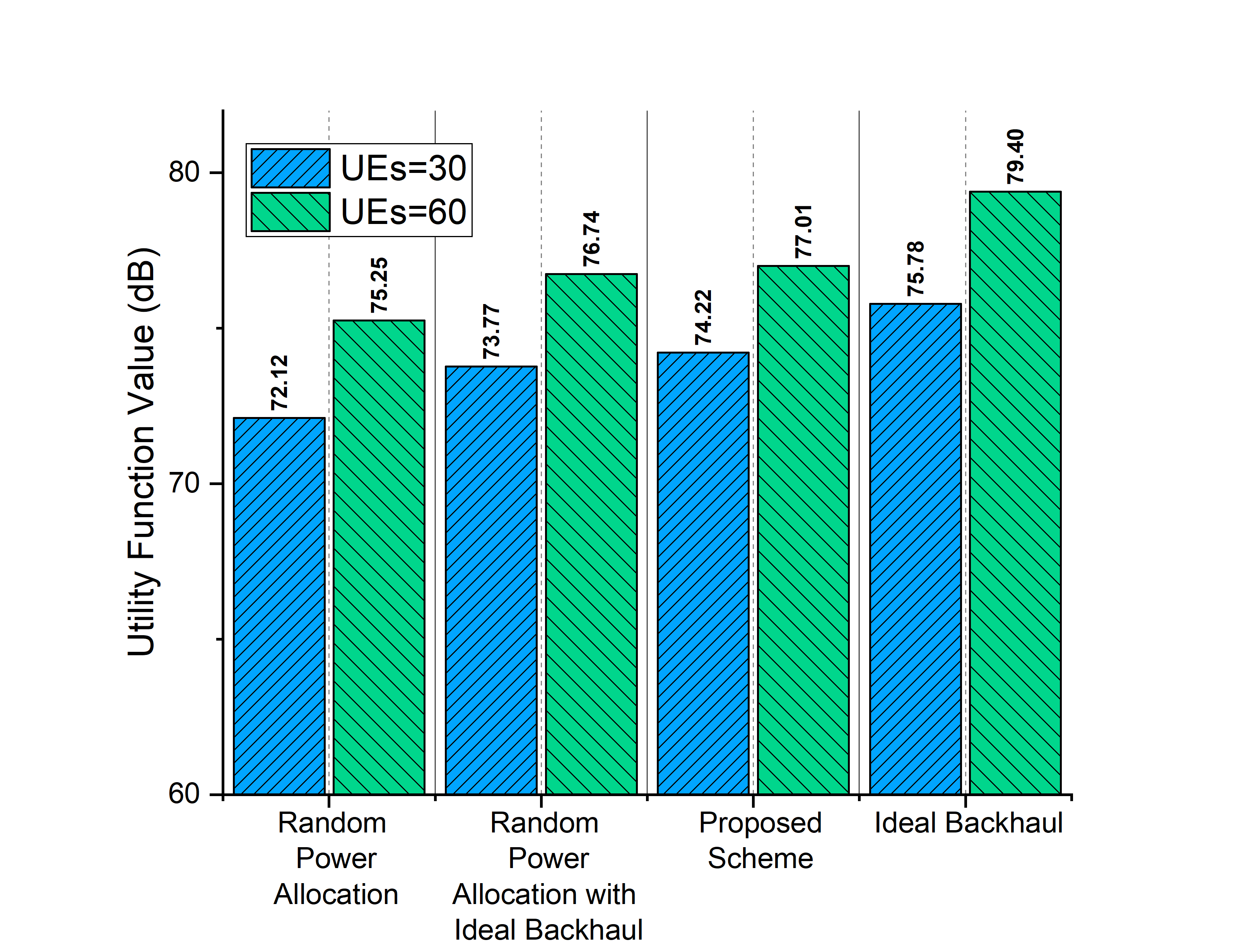}
	\caption{Comparison of Proposed With Others Benchmarks Schemes}
	\label{fig:R3}
\end{figure}
\begin{figure}[!t]
	\centering
	\includegraphics[width=0.8\linewidth]{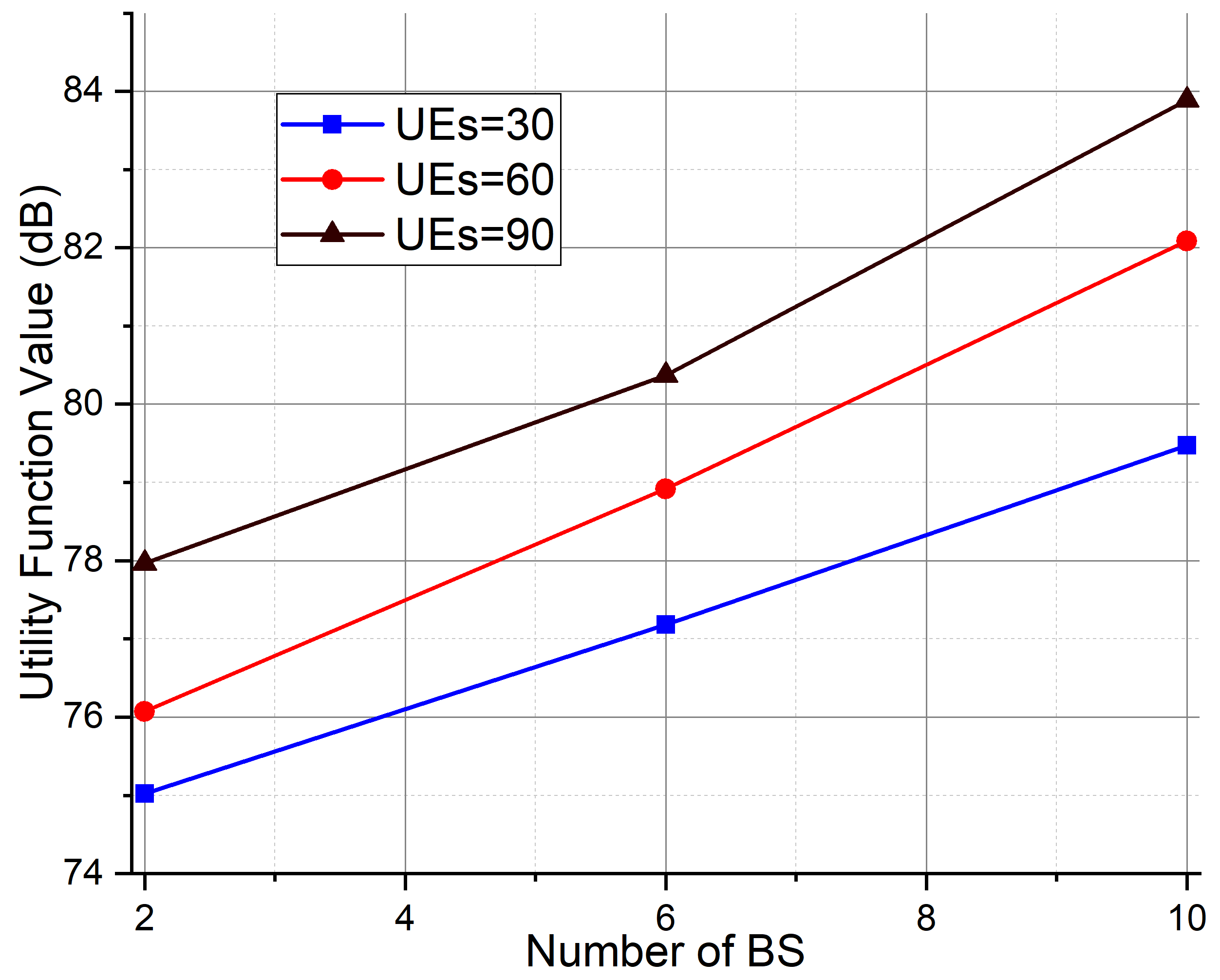}

	\caption{Utility Function value across the number of BSs.}
	\label{fig:R4}
\end{figure}
\begin{figure}[!t]
	\centering
	\includegraphics[width=0.8\linewidth]{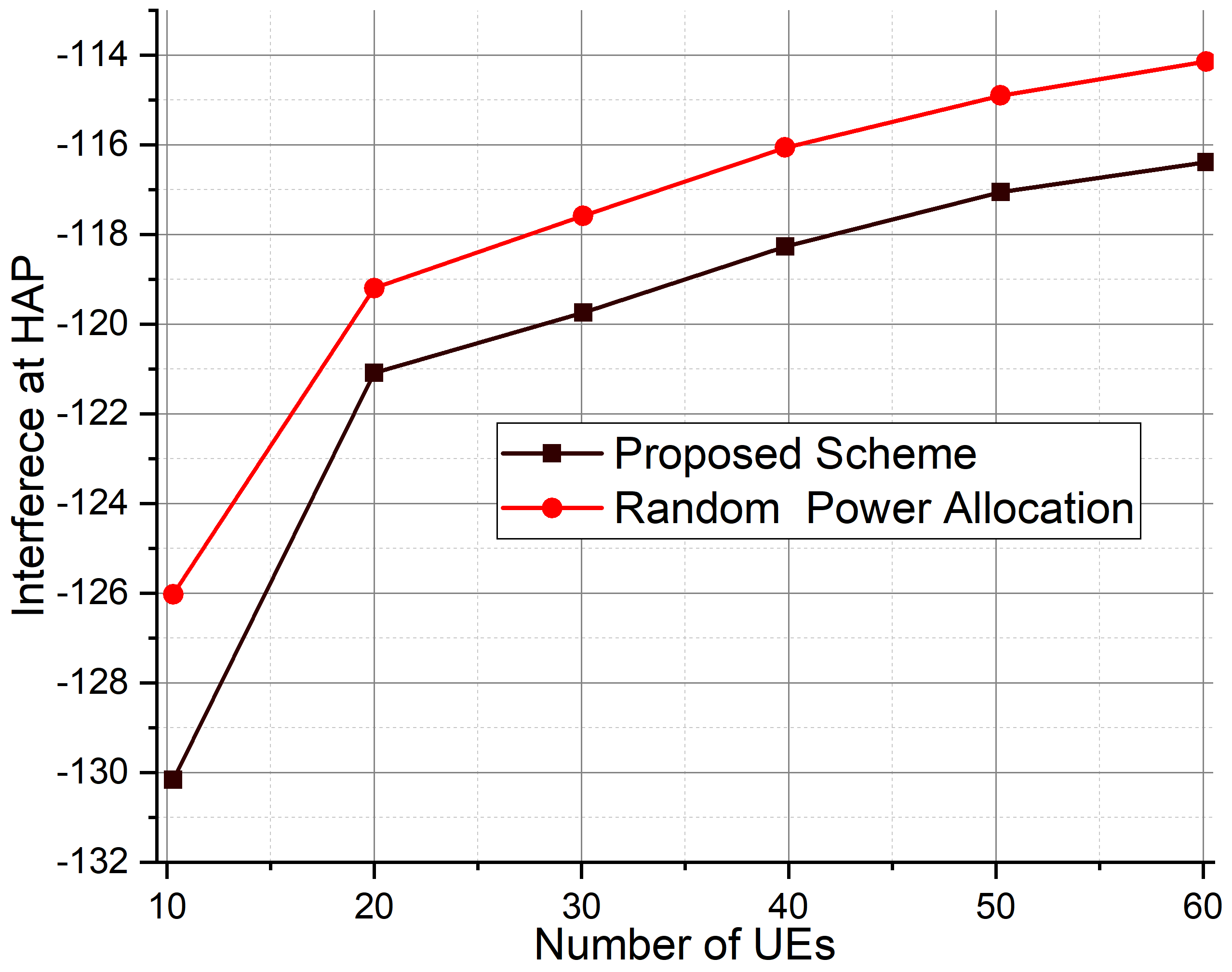}
	\caption{Utility Function value across the number of users}
	\label{fig:R5}
\end{figure}
\begin{figure}[!t]
	\centering
	\includegraphics[width=0.8\linewidth]{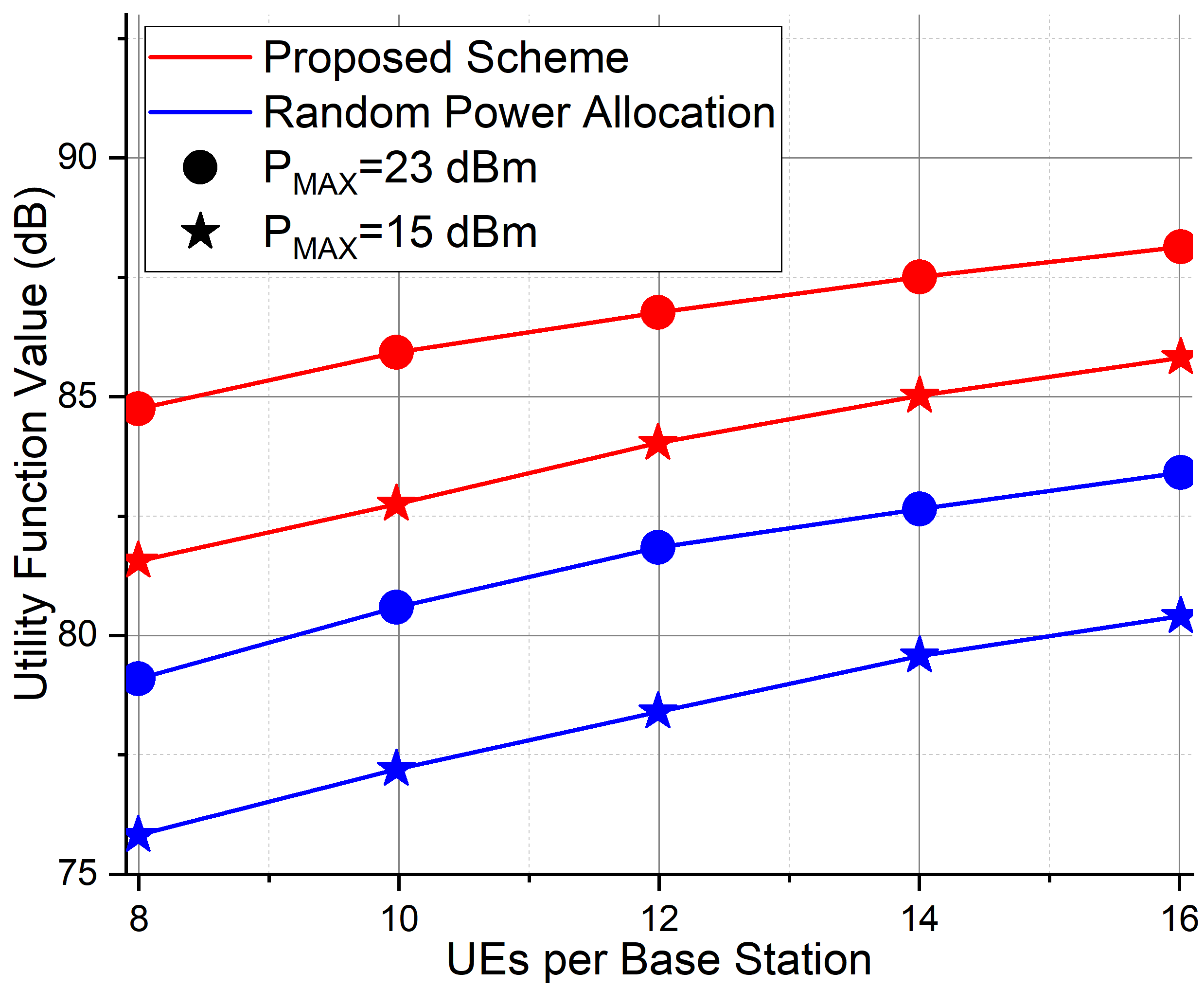}
	\caption{Utility Function value across users density}
	\label{fig:R6}
\end{figure}
\begin{figure}[!t]
	\centering
	\includegraphics[width=0.8\linewidth]{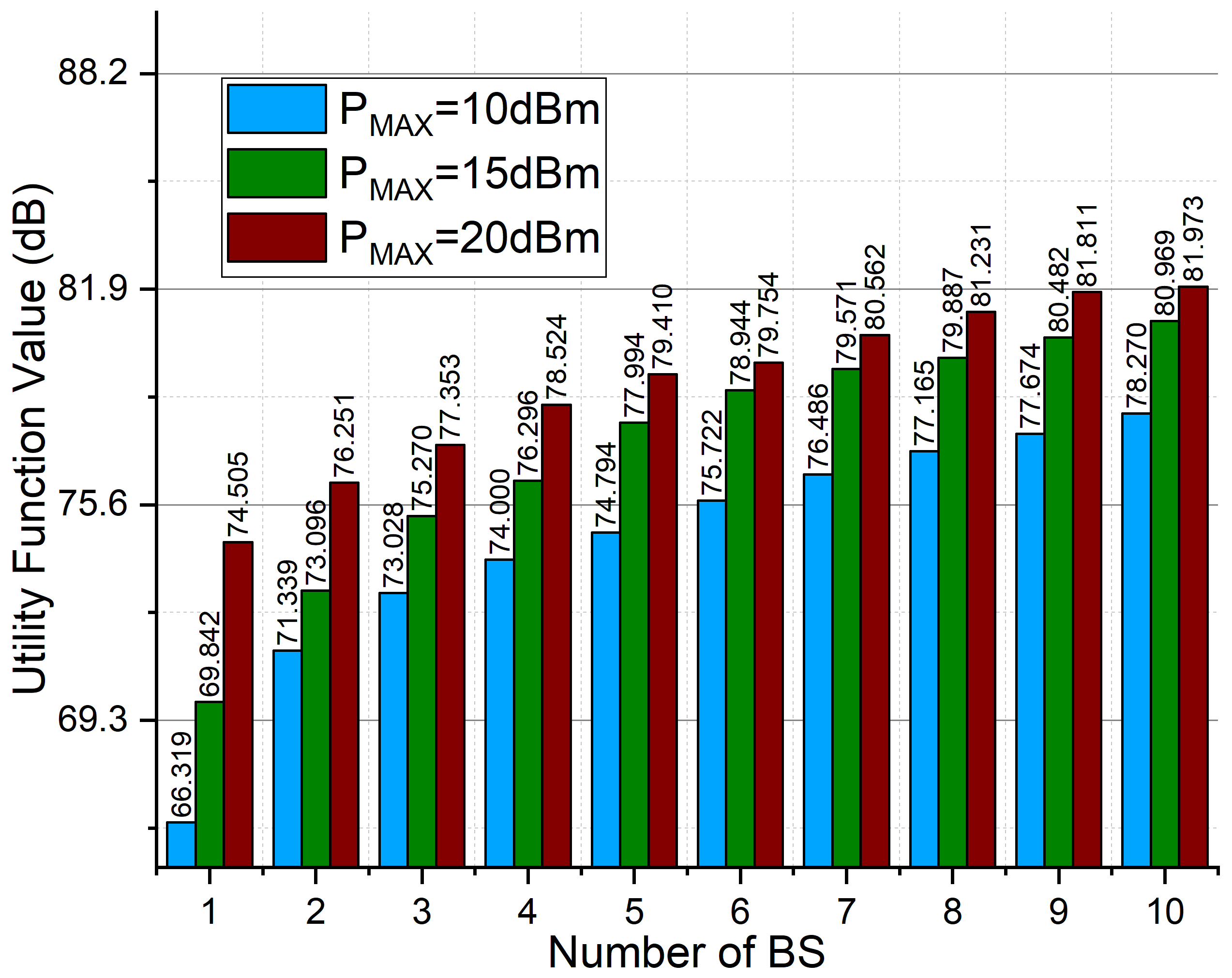}
	\caption{Utility Function value across Power Values}
	\label{fig:R7}
\end{figure}
\begin{figure}[!t]
	\centering
	\includegraphics[width=0.8\linewidth]{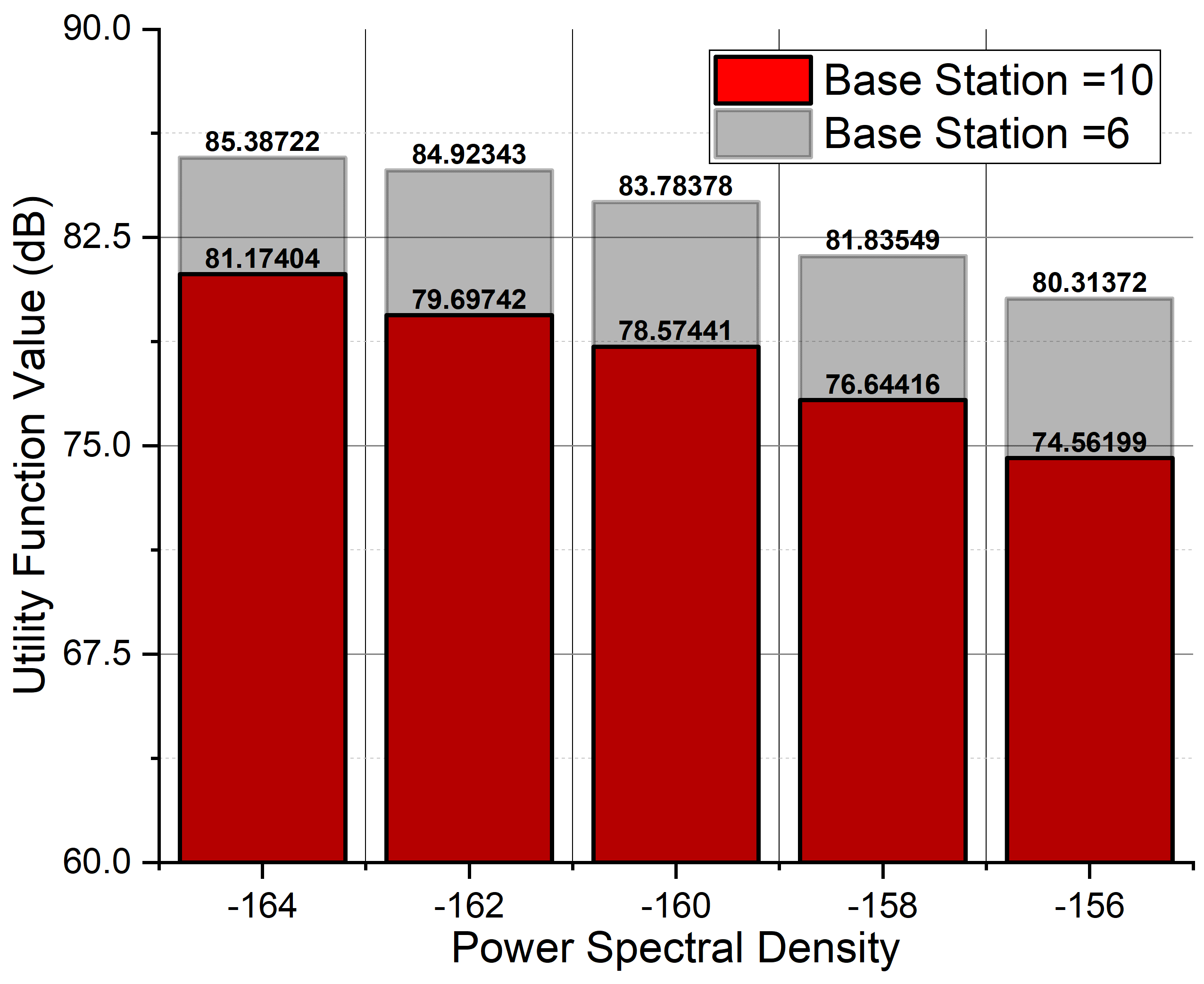}
	\caption{Utility Function value across Power Spectral Density}
	\label{fig:R8}
\end{figure}
\section{Result and Discussion}
This section presents the simulation results that demonstrate the effectiveness of our proposed algorithms in mitigating interference and maximizing the sum rate in a NOMA-enabled vehicular-aided HetNet. Mitigating interference is crucial for improving the performance of wireless communication networks, and the proposed algorithms aim to optimize resource allocation and interference management to achieve this goal. To mimic real-world conditions, we carefully chose the parameters used in the experiments. The HAP's altitude is fixed at $1000$ km, and the radius of each RSU is set to $50$ m. Each user device and RSU has a maximum transmission power capacity of $23$ dBm and $43$ dBm, respectively. The HetNet has a total system bandwidth of $20$ MHz and an additive white Gaussian noise power of $-174$ dBm/Hz. The weight factor in the proposed UE association approach is $0.99$, and there are $50$ VUEs and $5$ RSUs in the network. The Rayleigh and Rician fading models are used to model the terrestrial and satellite links, respectively.
\par
The results show that our proposed algorithms effectively mitigate interference and maximize the sum rate of the HetNet. Figure \ref{fig:R1} compares the performance of the proposed scheme with fixed user association algorithms by plotting their convergence. Each VUE's power is set to $23$ dBm, and the weight factor is set to $0.99$. The VUE association algorithm and the fixed association algorithm begin with rapid increases in their utility curves, as shown in the figure. However, the rate of increase slows as iterations continue until convergence. The proposed algorithm achieves convergence in $500$ to $600$ iterations, which is faster than the random swapping algorithm, which takes around $1000$ iterations. This indicates that our proposed UE association algorithm has low computational complexity and can effectively mitigate interference in the network, leading to a higher sum rate.
\par
The convergence of Algorithm \ref{algo:ALO2} is evaluated with respect to various VUE association schemes, as shown in Figure \ref{fig:R2}. The results demonstrate that the utility function value stabilizes, indicating that the transmission power allocation using the successive convex approximation-based algorithm has converged. Furthermore, the proposed algorithm outperforms the others by considering the utility function value as a performance metric.
\par
Simulation results in Figure \ref{fig:R3} show the impact of  VUE's on the utility function for various schemes. Similarly, the proposed scheme is compared to three other schemes: ideal backhaul, random power allocation, and random power allocation with ideal backhaul. The results demonstrate that as the number of VUE's increases, so does the utility function for all four schemes. At the same time, the proposed scheme provides the same epsilon results as the ideal backhaul approach. When combined with ideal backhaul schemes, it outperforms random power allocation and random power allocation schemes. With the Ideal Backhaul scheme, the optimization problem's tractable region is expanded, which improves the system's value in hybrid networks that combine HAPS and ground infrastructure.
\par
Figure \ref{fig:R4} depicts the relationship between the performance of the utility function and the VUE's for different RSU configurations. The graph reveals a positive correlation between the number of RSUs and the utility function performance. This can be attributed to the increased availability of candidate RSUs for each VUE when the number of VUE's is constant. The improved selection of candidate RSUs results in an enhancement in the overall system performance, as indicated by the utility metric.

\par
Similarly, the results in Figure \ref{fig:R5} analyze cross-tier interference impact on satellite networks. The proposed algorithm's performance is compared to the Random Power Allocation  approach, which utilizes the UE association scheme. Moreover, the results demonstrate that the system's overall performance improves with the UE increase. The proposed algorithm outperforms the random Power Allocation scheme by effectively reducing the cross-tier interference from BSs in the satellite network. In comparison, the same effect is achieved by considering the negative impact of cross-tier interference and regulating UE transmission power, ensuring optimal quality of service for satellite UEs.
\par
Similarly, to reveal the effectiveness of the proposed schemes, results are compared with the Random Power Allocation algorithm through an examination of the utility as a function of the UE density per RSU, as shown in Figure \ref{fig:R6}. The graph is based on a fixed number of $5$ RSUs. The results show that as the UE density per RSU increases, the utility function increases, thanks to the higher transmission power of UEs. Furthermore, the proposed algorithm achieves a significantly larger system utility value than the Random Power Allocation approach. This difference becomes more apparent with more UEs.
\par
Similarly, results in Figure \ref{fig:R7} show the relationship between the number of terrestrial RSUs and the efficiency of the system analyzed. The graph is based on an environment with $50$ UEs and varying maximum transmission power levels per UE: $[10, 25, 20]$ dBm. Results demonstrate that a rise in the number of RSUs leads to improved system efficiency, aligning with the trend seen when evaluating the UE density per RSU. Furthermore, elevating the maximum allowable transmission power for each UE enhances system efficiency by extending the feasible range of the power optimization problem.
\par
Figure \ref{fig:R8} depicts the effect of the AWGN power spectral density on system efficiency. This graph shows the impact of different levels of AWGN on system performance for cases with $10$ and $6$ terrestrial RSUs and a fixed number of 100 UEs. The graph shows that increasing the AWGN power spectral density reduces system efficiency. When the AWGN power spectral density is constant, the graph shows that increasing terrestrial RSUs improves system efficiency. This demonstrates the significant impact of the number of terrestrial RSUs on system efficiency in the presence of AWGN.
{\color{black} \section{Conclusion and Future Work}
In this study, we proposed a three-stage iterative resource optimization algorithm for a NOMA-based uplink caching heterogeneous network with an RSU-HAPs configuration. The proposed algorithm optimizes the resource allocation to improve the utility performance of the network while considering the QoS constraints of terrestrial VUEs and the backhaul constraints of backhaul HAPS. In the first stage of the algorithm, we developed an improved caching and swapping algorithm that incorporated preference relations to optimize the RSU-VUE association sub-problem. In the second stage, we derived a closed-form expression for the bandwidth allocation coefficient. Finally, the third stage utilized the successive convex approximation method to solve the non-convex power allocation sub-problem iteratively. The simulation results demonstrated that the proposed algorithm significantly improved the network's utility performance. The results also showed that increasing the number of terrestrial RSUs and the maximum allowable transmission power for each VUE led to improved system efficiency. Moreover, increasing the number of RSUs and decreasing the AWGN power spectral density improved system efficiency in the presence of AWGN. Overall, our proposed algorithm successfully optimized resource allocation and enhanced system efficiency while considering the constraints of a NOMA-based uplink caching heterogeneous network with an RSU-HAPs configuration.

For future research, we intend to expand this study to include more complex networks and explore other optimization aspects of such networks. Specifically, our focus will extend to areas such as minimizing latency and optimizing energy consumption. We aim to contribute to the ongoing development of advanced communication networks for smart cities and vehicular applications by addressing these vital aspects in our future work.}
\section*{Acknowledgment}
The authors extend their appreciation to the Deanship of Scientific Research at King Khalid University for funding this work through large group Research Project under grant number (RGP2/10/44). Princess Nourah bint Abdulrahman University Researchers Supporting Project number (PNURSP2024R384), Princess Nourah bint Abdulrahman University, Riyadh, Saudi Arabia. Research Supporting Project number (RSPD2024R787), King Saud University, Riyadh, Saudi Arabia. This study is supported via funding from Prince Sattam bin Abdulaziz University project number (PSAU/2023/R/1444).

\bibliographystyle{elsarticle-num}
\bibliography{egbib}

\end{document}